\documentclass[preprint,amsmath,amssymb,aps,a4,superscriptaddress]{revtex4-2}
\usepackage{graphicx}
\usepackage{bm}
\usepackage{txfonts}
\usepackage{hyperref}
\usepackage{subcaption}
\date{\today}

\newcommand{\be}{\begin{eqnarray}}
\newcommand{\ee}{\end{eqnarray}}

\newcommand{\bfp}{{\bf p}_{\perp}}

\usepackage{xcolor}
\begin{document}
	\title{Sub-leading twist transverse momentum dependent parton distributions in the light-front quark-diquark model}
	\author{Shubham Sharma}
	\email{s.sharma.hep@gmail.com}
	\affiliation{Department of Physics, Dr. B. R. Ambedkar National Institute of Technology, Jalandhar 144027, India }
	\author{Narinder Kumar}
	\email{narinderhep@gmail.com}
	\affiliation{Department of Physics, Dr. B. R. Ambedkar National Institute of Technology, Jalandhar 144027, India }
	\affiliation{Computational Theoretical High Energy Physics Lab, Department of Physics, Doaba College, Jalandhar 144004, India}
	\author{Harleen Dahiya}
	\email{dahiyah@nitj.ac.in}
	\affiliation{Department of Physics, Dr. B. R. Ambedkar National Institute of Technology, Jalandhar 144027, India }
	\date{\today}
	\begin{abstract}
		In this study, the T-even sub-leading twist transverse momentum dependent distributions (TMDs) of proton in the light-front quark-diquark model (LFQDM) have been investigated. We have derived the overlap form of the light-front wave functions (LFWFs) for the sub-leading twist proton TMDs by detangling the un-integrated quark-quark correlator for the semi-inclusive deep inelastic scattering (SIDIS).  We have obtained the explicit expressions of TMDs for both the cases of the diquark being a scalar or a vector and analysed their relationships with leading twist TMDs within the same model. Average transverse momenta and average square transverse momenta for the TMDs have been tabulated and compared with the results from the light-front bag model and the light-front constituent quark model (LFCQM). In addition to this, we have also compared our results for the PDF $e(x)$ with the recent CLAS collaboration results.
		%
		\par
		\vspace{0.1cm}
		\noindent{\it Keywords}: Transverse momentum dependent parton distributions of proton; T-even sub-leading twist TMDs; light-front quark-diquark model.
	\end{abstract}
	\maketitle
	%
	%
	\section{Introduction \label{secintro}}
	Being the generalization of \cite{Collins:2003fm,Collins:2007ph,Collins:1999dz,Hautmann:2007uw,PDF4LHCWorkingGroup:2022cjn} of the parton distribution functions (PDFs), the transverse momentum dependent parton distributions (TMDs) are the center of attention for particle physicists. The TMDs seem commending to strengthen our understanding about the structure of nucleon surpassing what we know  from the PDFs i.e., about the distribution of partons in the longitudinal momentum space. The TMDs  encrypt the three dimensional (3-D) structure along with the information on angular momentum of the nucleon
	and spin-orbit correlations \cite{Collins:1981uk,Ji:2004wu,Collins:2004nx,Cahn:1978se,Konig:1982uk,Chiappetta:1986yg,Collins:1984kg,Sivers:1989cc,Efremov:1992pe,Collins:1992kk,Collins:1993kq,Kotzinian:1994dv,Mulders:1995dh,Boer:1997nt,Boer:1997mf,Boer:1999mm,Bacchetta:1999kz,Brodsky:2002cx,Collins:2002kn,Belitsky:2002sm,Burkardt:2002ks,Pobylitsa:2003ty,Goeke:2005hb,Bacchetta:2006tn,Cherednikov:2007tw,Brodsky:2006hj,Avakian:2007xa,Miller:2007ae,Arnold:2008kf,Brodsky:2010vs,lattice-TMD}.
	\par
	The transverse momentum dependent fragmentation functions and TMDs are beneficial for a clear analysis of the leading twist observables in deep inelastic scattering (DIS) experiments \cite{Collins:1981uk,Ji:2004wu,Collins:2004nx}, for which we have data from hadron production in $e^+e^-$ annihilations \cite{Abe:2005zx,Ogawa:2006bm,Seidl:2008xc,Vossen:2009xz} and the Drell-Yan processes \cite{Falciano:1986wk,Conway:1989fs,Zhu:2006gx}. Further, semi-inclusive deep inelastic scattering (SIDIS) processes \cite{Arneodo:1986cf,Airapetian:1999tv,Avakian:2003pk,Airapetian:2004tw,Alexakhin:2005iw,Gregor:2005qv,Ageev:2006da,Airapetian:2005jc,Kotzinian:2007uv,Diefenthaler:2005gx,Airapetian:2008sk,Osipenko:2008rv,Giordano:2009hi,Gohn:2009,Airapetian:2009jy} give extensive knowledge on the partons structure, including sub-leading twist effects. In the higher order quantum chromodynamics (QCD) computations \cite{Gehrmann:2014yya,Echevarria:2015byo,Echevarria:2016scs,Li:2016ctv,Vladimirov:2016dll,Gutierrez-Reyes:2017glx,Gutierrez-Reyes:2018iod,Luo:2019hmp,Luo:2019szz,Ebert:2020yqt} and phenomenological research \cite{Efremov:2004tp,Anselmino:2005nn,Vogelsang:2005cs,Collins:2005ie,Collins:2005rq,Anselmino:2007fs,Anselmino:2013vqa,Signori:2013mda,Anselmino:2013lza,Kang:2014zza,Kang:2015msa,Kang:2017btw,Cammarota:2020qcw,Lefky:2014eia}, tremendous progress has been made. In addition, various facets of TMD physics have been examined in depth \cite{Collins:2003fm,DAlesio:2007bjf,Barone:2010zz,Aidala:2012mv,Avakian:2019drf,Anselmino:2020vlp}. The transverse structure of the nucleon has been examined with the aid of TMDs in the SIDIS processes, and this is one of the driving forces behind the development of the Electron-Ion Collider (EIC) at  Brookhaven National Laboratory (BNL) \cite{Accardi:2012qut,EIC2103.05419,EIC2203.13199,EIC2203.13923}. The cross-sections of the above said processes are great source of information on the nature of fundamental interactions. The factorization theorem \cite{eche1,eche2,eche3,collins} for these cross-sections pointed out that the processes like SIDIS and DY encodes all the non-perturbative effects in the form of TMDs and can be included in experiments running at different energies by solving the evolution equations.  TMDs evolution is usually regulated by the Collins-Soper-Sterman equation and it depends on perturbative inputs from the transverse momentum resummation\cite{Catani:2000vq,Bozzi:2005wk,Bozzi:2008bb}.  
	\par
	Calculating the TMDs in full QCD is an overwhelming task due to their non-perturbative character, however, the TMDs have been computed using a variety of QCD-inspired models. Distributions of hadrons (TMDs, PDFs, form factor distributions, etc.) have been investigated in quark-target \cite{Kundu:2001pk,Meissner:2007rx,Mukherjee:2009uy,Mukherjee:2010iw,Xu:2019xhk}, holographic \cite{Maji:2017wwd,Lyubovitskij:2020otz}, Nambu–Jona-Lasinio \cite{Matevosyan:2011vj}, Valon \cite{Yazdi:2014zaa}, light-front constituent quark (LFCQM) \cite{Pasquini:2008ax,Pasquini:2010af,Lorce:2011dv,Boffi:2009sh,Pasquini:2011tk,Lorce:2014hxa,Kofler:2017uzq,Pasquini:2018oyz,Rodini:2019ktv}, quark-diquark \cite{Jakob:1997wg,Gamberg:2007wm,Cloet:2007em,Bacchetta:2008af,She:2009jq,Lu:2012gu,Maji:2015vsa,Maji:2016yqo,Maji:2017bcz}, covariant parton model (CPM) \cite{Bastami:2020rxn}, chiral quark soliton \cite{Diakonov:1996sr,Diakonov:1997vc,Gamberg:1998vg,Pobylitsa:1998tk,Goeke:2000wv,Wakamatsu:2000fd,Schweitzer:2001sr,Schweitzer:2003uy,Wakamatsu:2003uu,Ohnishi:2003mf,Cebulla:2007ej,Wakamatsu:2009fn,
		Schweitzer:2012hh} and bag models \cite{Jaffe:1991ra,Yuan:2003wk,Courtoy:2008vi,Avakian:2008dz, Courtoy:2008dn,Avakian:2010br}. In some cases, model-independent lattice QCD computations have also been performed \cite{lattice-TMD,Musch:2010ka,Musch:2011er,
		Chen:2016utp,Alexandrou:2016jqi,Yoon:2017qzo,Orginos:2017kos,Joo:2019jct}.  
	The model based calculations of the TMDs are actually non-perturbative and therefore does not describe the complete distribution.
	\par Sub-leading and higher twist distributions for hadrons have been discussed in  Refs.  \cite{Avakian:2010br,Jakob:1997wg,Lorce:2014hxa,Pasquini:2018oyz,Kundu:2001pk,Mukherjee:2010iw,Lorce:2014hxa,PhysRevLett.67.552,SIGNAL1997415,PhysRevD.95.074017,ELLIS19821,ELLIS198329,QIU1991105,QIU1991137,PhysRevD.83.054010,liu21,Sharma:2023azu}. Specifically, sub-leading twist T-even unpolarized TMDs  $e^{\nu}(x, {\bf p_\perp^2})$ and $f^{\perp\nu}(x, {\bf p_\perp^2})$  have been discussed for spin$-\frac{1}{2}$ baryons in  Refs. \cite{Avakian:2010br,Jakob:1997wg,Lorce:2014hxa,Pasquini:2018oyz,Mukherjee:2010iw}. Study of sub-leading twist TMDs for these baryons have been performed in bag model \cite{Avakian:2010br}, spectator model \cite{Jakob:1997wg,liu21} and CPM \cite{Bastami:2020rxn}. In light of the above developments, extending the study of sub-leading twist TMDs in detail i.e., representation of their variation with the TMD variables, average transverse momentum values, relations with leading twist variables and analysis of model independent or dependent relations among them becomes desirable. 
	\par 
	For a two-body bound state, the light-front AdS/QCD  has predicted many intriguing nucleon features, particularly the form of its wave function.
	\cite{BT,EPJC77,PRD83,PRD89,PRD91}. It is in agreement with the quark counting rule \cite{Maji:2016yqo} and the Drell-Yan-West relation \cite{DY70,West70}. The major goal of AdS/QCD is to combine two theories—Anti-de Sitter space (AdS) in string theory and QCD, which at first glance appear to be unconnected. According to the AdS/QCD theory, some QCD observables can be mapped onto a dual description in AdS space, where computations are easier to perform. In the light-front quark-diquark model (LFQDM), the proton is represented as a combination of an active quark and a diquark spectator with a specific mass \cite{Chakrabarti:2019wjx}. Basically, the light-front wave functions (LFWFs) derived from the predictions of AdS/QCD include contributions from the scalar ($S=0$) and axial vector ($S=1$) diquarks and have a $SU(4)$ spin-flavor structure \cite{Maji:2016yqo}.
	The LFWFs of the LFQDM provide detailed insights into the probability amplitudes of discovering particular quark and diquark configurations within the proton. The dynamics of confined states, such as confinement and quark-diquark interactions, naturally incorporated into the LFQDM. A fully relativistic treatment of hadron physics provided by the LFQDM developed within the context of light-front dynamics. In LFQDM, PDFs have been evolved from $\mu^2=0.09~{\rm GeV}^2$ to any arbitrary scale (up to $\mu^2=10^4~{\rm GeV}^2$), therefore we can calculate distributions at any given scale.
	Further, in LFQDM the transversity and helicity PDFs have been analyzed and they have been found to be in good agreement with the experimental results. It has been also demonstrated that the Soffer bound is in line with the existing data and the transversity TMD satisfies it. The experimental values of axial and tensor charges have also been replicated. 
	%
	For the quarks in a proton, the generalized parton distributions (GPDs) have been investigated in the LFQDM in both position and momentum spaces \cite{Mondal:2015uha}. The outcomes have been compared to the soft-wall AdS/QCD model for the proton GPDs with zero skewness. The GPDs for non-zero skewness have also been computed. The GPDs show a diffraction pattern in longitudinal position space, which has also been shown by other models \cite{Mondal:2015uha}. Comparative study of the nucleon charge and anomalous magnetization density in the transverse plane has been carried out \cite{Mondal:2015uha}. Calculations on flavor decomposition of the transverse densities and form factors have also been performed \cite{Mondal:2015uha}.
	In LFQDM, the leading twist T-odd quark TMDs of the proton have also been obtained, specifically the Sivers function $f_{1T}^{\perp q}(x, {\bf p_\perp^2})$ and the Boer-Mulders function $h_{1}^{\perp q}(x, {\bf p_\perp^2})$ \cite{Gurjar:2022rcl}. The generalized Sivers and Boer-Mulders shifts have also been compared to the known lattice QCD simulations and it has been proved that the SIDIS spin asymmetries related to these T-odd TMDs are consistent with COMPASS and HERMES findings \cite{Gurjar:2022rcl}.\\
	Leading twist T-even TMDs, comprising both the scalar and vector diquarks, have been explored in LFQDM \cite{Maji:2017bcz}. Common diquark model inequalities are found to be satisfied in LFQDM as well. Even though $x$-$p_\perp$ factorization is not observed, unlike other phenomenological models for the TMD $f_1^\nu(x,{\bf p_\perp^2})$, the numerical study of TMDs in LFQDM is consistent with the phenomenological ansatz. In LFQDM, the transverse shape of the proton has been depicted progressively. LFQDM has demonstrated, for leading twist TMDs, the relationship between quark densities and the first moments in $x$ of TMDs $f_1^\nu(x,{\bf p_\perp^2})$ and $g_{1T}^\nu(x,{\bf p_\perp^2})$ for various polarizations of quarks and the parent proton. PDFs can be derived successively by $p_\perp$-integration of the TMDs $f_1^\nu(x,{\bf p_\perp^2}),~ h_1^\nu(x,{\bf p_\perp^2})$ and $g_{1L}^\nu(x,{\bf p_\perp^2})$, however, there are no collinear interpretations for the other TMDs. Some specific ratios, such as $g_{1T}^\nu(x)/h_{1L}^\nu(x)$ and $h_{1T}^\nu(x)/f_1^\nu(x)$, do not depend on the evolution scale $\mu$, when integrated TMDs are processed through DGLAP evolution at high scales in LFQDM. The relations between TMDs and GPDs in a LFQDM have been studied and many of the relations which we have obtained seem to have a similar structure in several models \cite{Gurjar:2021dyv}. The relation between the  GPD $E_q$ and the Sivers function can be derived in terms of a lensing function in LFQDM \cite{Gurjar:2021dyv}. The orbital angular momentum of quarks is computed and compared to the results of other similar models \cite{Gurjar:2021dyv}. \\
	In LFQDM, the gravitational form factors (GFFs) and the mechanical properties, i.e. the mechanical radius, shear forces within the proton, and pressure distributions  have been also investigated \cite{Chakrabarti:2020kdc}. The results for GFFs, $A(Q^2)$ and $B(Q^2)$, are in good agreement  with the lattice QCD, whereas the qualitative behavior of the D-term form factor is on the same page with the obtained data from the JLab deeply virtual Compton scattering (DVCS) experiments, the predictions of various phenomenological models and the lattice QCD \cite{Chakrabarti:2020kdc}. The distributions of shear force and pressure are also in agreement with the outcomes of other models \cite{Chakrabarti:2020kdc}.

	%
	%
	
	\par Following the successes of the LFQDM, we have investigated the sub-leading twist T-even TMDs for proton in this work. Sub-leading twist TMDs are usually decomposed into three contributions i.e., singularity term which relates the pion-nucleon sigma term and QCD vaccum structure, the tilde term which is related with genuine quark-gluon-quark contribution and third term is quark mass term. Since LFQDM does not have any contributions from gluon part, therefore we have constrained ourselves to consider the only mass term. This is also refered to as Wandzura-Wilczek approximation \cite{Wandzura:1977qf}. In Ref. \cite{1202.0700}, T-odd TMDs are calculated using the quark-gluon-quark correlation function. Using the diquark model, as a lowest approximation, gauge-links were ignored and it shows that T-odd distributions are non-vanishing. For SIDIS, we have deciphered the un-integrated quark-quark correlator and derived the overlap form of the TMDs for an appropriate proton polarization. Their exact expressions are provided for both the scalar and vector cases of diquark.  The 2-D and 3-D variation of sub-leading twist T-even TMDs for the $u$ and $d$ quarks, with longitudinal momentum fraction $x$ and transverse momentum ${\bf p_\perp^2}$, have been explored.  We have expressed our results of the sub-leading twist T-even TMDs in the form of  available leading twist TMDs \cite{Maji:2017bcz}. We have tabulated and compared the results of average transverse momenta and average transverse momenta square for our sub-leading twist T-even TMDs with the results of the LFCQM \cite{Lorce:2014hxa} and bag model \cite{Avakian:2010br}. We have also presented the PDFs $\big(x e^{\nu}(x),~ x f^{\perp\nu}(x),~x g_{L}^{\perp\nu}(x), x g_{T}^{'\nu}(x),~x g_{T}^{\perp\nu}(x),$ $x h_{L}^{\nu}(x),~x h_{T}^{\nu}(x)$ and $ x h_{T}^{\perp\nu}(x)\big)$ generated by integrating the TMDs over the transverse momentum of quark. In addition to this, we have compared our result for the PDF $e(x)$ with the available experimental data from CLAS collaboration.
	
	\par Our paper is organized as follows. In Sec. \ref{secmodel}, we have covered the quark-diquark model's characteristics and model parameters. In Sec. \ref{sectmdsp}, projections of the sub-leading twist quark TMDs have been presented in the form of a quark-quark correlator. In section \ref{sectmdsp}, we have discussed the sub-leading twist TMDs and their correlator. In Sec. \ref{secresults}, we have presented the results of the sub-leading twist T-even TMDs in the overlap form of LFWFs with their explicit representations. We have interpreted TMDs in Sec. \ref{secdiscussion} with the aid of 2D and 3D Plots. The link between sub-leading and leading twist T-even TMDs has been discussed. In the same section, we have analyzed the results of average transverse momenta and average transverse momenta square for sub-leading twist T-even TMDs and compared them to previous findings of LFCQM and bag model. We have also discussed their PDFs in this section. 
	In Sec. \ref{pheno}, we have compared our results for PDF $e(x)$ with the CLAS data.
	Finally, Sec. \ref{seccon} concludes with a summary of the outcomes. 
	%
	%
	\section{Light-Front Quark-Diquark Model (LFQDM) \label{secmodel}}
	In the LFQDM \cite{Maji:2016yqo}, the proton is described as an aggregate of active quark and a diquark spectator of definite mass \cite{Chakrabarti:2019wjx}.
	The proton has spin-flavor $SU(4)$ structure and it has been stated as a composite of isoscalar-scalar diquark singlet $|u~ S^0\rangle$, isoscalar-vector diquark $|u~ A^0\rangle$ and isovector-vector diquark $|d~ A^1\rangle$ states as \cite{Jakob:1997wg,Bacchetta:2008af}
	\begin{equation}
		|P; \pm\rangle = C_S|u~ S^0\rangle^\pm + C_V|u~ A^0\rangle^\pm + C_{VV}|d~ A^1\rangle^\pm. \label{PS_state}
	\end{equation}
	Here, $S$ and $A=V,VV$ have been used to denote the scalar and vector diquark respectively. Their isospin has been represented by the superscripts on them. The coefficients $C_{i}$ of scalar and vector diquarks have been established in Ref. \cite{Maji:2016yqo} and given as
	\begin{equation}
		\begin{aligned}
			C_{S}^{2} &=1.3872, \\
			C_{V}^{2} &=0.6128, \\
			C_{V V}^{2} &=1.
		\end{aligned}
		\label{Eq3d1}
	\end{equation}
	We have used the light-cone convention $z^\pm=z^0 \pm z^3$ and the frame is selected such that the proton's transverse momentum does not exist i.e., $P \equiv \big(P^+,\frac{M^2}{P^+},\textbf{0}_\perp\big)$. The momentum of the struck quark ($p$) and diquark ($P_X$) are
	\be
	p &&\equiv \bigg(xP^+, \frac{p^2+|\bfp|^2}{xP^+},\bfp \bigg),\\
	P_X &&\equiv \bigg((1-x)P^+,P^-_X,-\bfp\bigg).
	\ee
	The longitudinal momentum fraction acquired by the struck quark has been denoted by $x=p^+/P^+$. The expansion of Fock-state in the case of two particle for $J^z =\pm1/2$ for the scalar diquark  can be expressed as
	\be
	|u~ S\rangle^\pm & =& \int \frac{dx~ d^2\bfp}{2(2\pi)^3\sqrt{x(1-x)}} \Bigg[ \psi^{\pm(u)}_{+}(x,\bfp)\bigg|+\frac{1}{2}~s; xP^+,\bfp\bigg\rangle \nonumber \\
	&+& \psi^{\pm(u)}_{-}(x,\bfp) \bigg|-\frac{1}{2}~s; xP^+,\bfp\bigg\rangle\Bigg],\label{fockSD}
	\ee
	where $\nu ~(=u,d)$ is the flavor index and $|\lambda_q~\lambda_S; xP^+,\bfp\rangle$ illustrates the state of two particles with the helicity of a struck quark as $\lambda_q$ and the helicity of a scalar diquark as $\lambda_S$. Here $\lambda_S=s$ represents the helicity of the singlet spin-0 diquark. Following Ref. \cite{Maji:2017bcz}, the LFWFs for the scalar diquark are given in Table \ref{tab_LFWF_s}. The generic ansatz of LFWFs $\varphi^{(\nu)}_{i}=\varphi^{(\nu)}_{i}(x,\bfp)$ has been adopted from the soft-wall AdS/QCD prediction \cite{BT,majiref27}. 
	
	\begin{table}[h]
		\centering 
		\begin{tabular}{ |p{1.2cm}|p{1.4cm}|p{1.2cm}|p{1.8cm} p{3.5cm}|p{1.8cm} p{3.5cm}|}
			\hline
			~~S No.~~&~~$\lambda_q$~~&~~$\lambda_S$~~&\multicolumn{2}{c|}{LFWFs for $J^z=+1/2$} & \multicolumn{2}{c|}{LFWFs for $J^z=-1/2$}\\
			\hline
			~~$1$~~&~~$+1/2$~~&~~$~~0$~~&~~$\psi^{+(\color{red}{\nu})}_{+}(x,\bfp)$~&~~$=~N_S~ \varphi^{(\nu)}_{1}$~~&~~$\psi^{-(\color{red}{\nu})}_{+}(x,\bfp)$~&~~$=~N_S \bigg(\frac{p^1-ip^2}{xM}\bigg)~ \varphi^{(\nu)}_{2}$~~  \\
			~~$2$~~&~~$-1/2$~~&~~$~~0$~~&~~$\psi^{+(\color{red}{\nu})}_{-}(x,\bfp)$~&~~$=~-N_S\bigg(\frac{p^1+ip^2}{xM} \bigg)~ \varphi^{(\nu)}_{2}$~~&~~$\psi^{-(\color{red}{\nu})}_{-}(x,\bfp)$~&~~$=~N_S~ \varphi^{(\nu)}_{1}$~~   \\
			\hline
		\end{tabular}
		\caption{The LFWFs for the scalar diquark for the case when $J^z=\pm1/2$, for different values of helicities of struck quark $\lambda_q$ and $\lambda_S=0$. $N_S$ is the normalization constant.}
		\label{tab_LFWF_s} 
	\end{table}

	Similarly, the Fock-state expansion for the vector diquark in the situation of two particles is given as \cite{majiref25}
	\be
	|\nu~ A \rangle^\pm & =& \int \frac{dx~ d^2\bfp}{2(2\pi)^3\sqrt{x(1-x)}} \Bigg[ \psi^{\pm(\nu)}_{++}(x,\bfp)\bigg|+\frac{1}{2}~+1; xP^+,\bfp\bigg\rangle \nonumber\\
	&+& \psi^{\pm(\nu)}_{-+}(x,\bfp)\bigg|-\frac{1}{2}~+1; xP^+,\bfp\bigg\rangle +\psi^{\pm(\nu)}_{+0}(x,\bfp)\bigg|+\frac{1}{2}~0; xP^+,\bfp\bigg\rangle \nonumber \\
	&+& \psi^{\pm(\nu)}_{-0}(x,\bfp)\bigg|-\frac{1}{2}~0; xP^+,\bfp\bigg\rangle + \psi^{\pm(\nu)}_{+-}(x,\bfp)\bigg|+\frac{1}{2}~-1; xP^+,\bfp\bigg\rangle \nonumber\\
	&+& \psi^{\pm(\nu)}_{--}(x,\bfp)\bigg|-\frac{1}{2}~-1; xP^+,\bfp\bigg\rangle  \Bigg].\label{fockVD}
	\ee
	Here $|\lambda_q~\lambda_D; xP^+,\bfp\rangle$ is the state of two particles with quark helicity of $\lambda_q=\pm\frac{1}{2}$ and vector diquark helicity of $\lambda_D=\pm 1,0$ (triplet). In Table \ref{tab_LFWF_v}, the LFWFs for the vector diquark have been listed when  $J^z=\pm1/2$ \cite{Maji:2017bcz}.
	\begin{table}[h]
		\centering 
		\begin{tabular}{ |p{1.2cm}|p{1.4cm}|p{1.2cm}|p{1.8cm} p{4.0cm}|p{1.8cm} p{4.0cm}|  }
			\hline
			~~S No.~~&~~$\lambda_q$~~&~~$\lambda_D$~~&\multicolumn{2}{c|}{LFWFs for $J^z=+1/2$} & \multicolumn{2}{c|}{LFWFs for $J^z=-1/2$}\\
			\hline
			~~$1$~~&~~$+1/2$~~&~~$+1$~~&~~$\psi^{+(\nu)}_{+~+}(x,\bfp)$~&~~$=~~N^{(\nu)}_1 \sqrt{\frac{2}{3}} \bigg(\frac{p^1-ip^2}{xM}\bigg)~  \varphi^{(\nu)}_{2}$~~&~~$\psi^{-(\nu)}_{+~+}(x,\bfp)$~&~~$=~~0$~~  \\
			~~$2$~~&~~$-1/2$~~&~~$+1$~~&~~$\psi^{+(\nu)}_{-~+}(x,\bfp)$~&~~$=~~N^{(\nu)}_1 \sqrt{\frac{2}{3}}~ \varphi^{(\nu)}_{1}$~~&~~$\psi^{-(\nu)}_{-~+}(x,\bfp)$~&~~$=~~0$~~   \\
			~~$3$~~&~~$+1/2$~~&~~$~~0$~~&~~$\psi^{+(\nu)}_{+~0}(x,\bfp)$~&~~$=~~-N^{(\nu)}_0 \sqrt{\frac{1}{3}}~  \varphi^{(\nu)}_{1}$~~&~~$\psi^{-(\nu)}_{+~0}(x,\bfp)$~&~~$=~~N^{(\nu)}_0 \sqrt{\frac{1}{3}} \bigg( \frac{p^1-ip^2}{xM} \bigg)~  \varphi^{(\nu)}_{2}$~~   \\
			~~$4$~~&~~$-1/2$~~&~~$~~0$~~&~~$\psi^{+(\nu)}_{-~0}(x,\bfp)$~&~~$=~~N^{(\nu)}_0 \sqrt{\frac{1}{3}} \bigg(\frac{p^1+ip^2}{xM} \bigg)~ \varphi^{(\nu)}_{2}$~~&~~$\psi^{-(\nu)}_{-~0}(x,\bfp)$~&~~$=~~N^{(\nu)}_0\sqrt{\frac{1}{3}}~  \varphi^{(\nu)}_{1}$~~   \\
			~~$5$~~&~~$+1/2$~~&~~$-1$~~&~~$\psi^{+(\nu)}_{+~-}(x,\bfp)$~&~~$=~~0$~~&~~$\psi^{-(\nu)}_{+~-}(x,\bfp)$~&~~$=~~- N^{(\nu)}_1 \sqrt{\frac{2}{3}}~  \varphi^{(\nu)}_{1}$~~   \\
			~~$6$~~&~~$-1/2$~~&~~$-1$~~&~~$\psi^{+(\nu)}_{-~-}(x,\bfp)$~&~~$=~~0$~~&~~$\psi^{-(\nu)}_{-~-}(x,\bfp)$~&~~$=~~N^{(\nu)}_1 \sqrt{\frac{2}{3}} \bigg(\frac{p^1+ip^2}{xM}\bigg)~  \varphi^{(\nu)}_{2}$~~   \\
			\hline
		\end{tabular}
		\caption{The LFWFs for the vector diquark for the case when $J^z=\pm1/2$, for different values of helicities of struck quark $\lambda_q$ and vector diquark $\lambda_D$. $N^{(\nu)}_0$, $N^{(\nu)}_1$ are the normalization constants.}
		\label{tab_LFWF_v} 
	\end{table}

	The generic ansatz of LFWFs $\varphi^{(\nu)}_{i}=\varphi^{(\nu)}_{i}(x,\bfp)$ shown in Table \ref{tab_LFWF_s} and \ref{tab_LFWF_v} is derived from the soft-wall AdS/QCD prediction \cite{BT,majiref27} and the parameters $a^\nu_i,~b^\nu_i$ and $\delta^\nu$ are established as follows \cite{Maji:2017bcz}
	\be
	\varphi_i^{(\nu)}(x,\bfp)=\frac{4\pi}{\kappa}\sqrt{\frac{\log(1/x)}{1-x}}x^{a_i^\nu}(1-x)^{b_i^\nu}\exp\Bigg[-\delta^\nu\frac{\bfp^2}{2\kappa^2}\frac{\log(1/x)}{(1-x)^2}\bigg].
	\label{LFWF_phi}
	\ee
	The wave functions $\varphi_i^\nu ~(i=1,2)$ are not symmetric under the exchange  $x \rightarrow 1-x$ and this asymmetry exists even at AdS/QCD limit $a_i^\nu=b_i^\nu=0$  and $\delta^\nu=1.0$. This asymmetry is arising from the matching of matrix elements of bare electromagnetic current between the dressed LFWF in the light-front QCD \cite{Gutsche:2013zia,Gutsche:2014yea}) and of the dressed electromagnetic current between the hadronic wave functions in AdS/QCD \cite{Brodsky:2007hb}-\cite{Gutsche:2013vb}.

	The parameters $a_i^{\nu}$ and $b_i^{\nu}$, which occur in Eq. \eqref{LFWF_phi}, have been fitted to the model scale $\mu_0=0.313{\ \rm GeV}$ using the Dirac and Pauli form factor data \cite{Maji:2016yqo,majiref16,majiref17}. At the model scale, the parameter $\delta^{\nu}$ is assumed to be one for both the $u$ and $d$ quarks \cite{Maji:2016yqo}. However, model parameters for $u$ and $d$ quarks are listed in Table \ref{tab_par}.
	\begin{table}[h]
		\centering 
		\begin{tabular}{ |p{1.4cm}|c|c|c|c|p{1.4cm}| }
			\hline
			~~$\nu$~~&~~$a_1^{\nu}$~~&~~$b_1^{\nu}$~~&~~$a_2^{\nu}$~~&~~$b_2^{\nu}$~~&~~$\delta^{\nu}$~~  \\
			\hline
			~~$u$~~&~~$0.280\pm 0.001$~~&~~$0.1716 \pm 0.0051$~~&~~$0.84 \pm 0.02$~~&~~$0.2284 \pm 0.0035$~~&~~$1.0$~~  \\
			~~$d$~~&~~$0.5850 \pm 0.0003$~~&~~$0.7000 \pm 0.0002$~~&~~$0.9434^{+0.0017}_{-0.0013}$~~&~~$0.64^{+0.0082}_{-0.0022}$~~&~~$1.0$~~    \\
			\hline
		\end{tabular}
		\caption{Values of model parameters corresponding to $u$ and $d$ quarks appearing in Eq. {\eqref{LFWF_phi}}.}
		\label{tab_par} 
	\end{table}	
	Apart from these, the normalization constants $N_{i}^{2}$ in Table {\ref{tab_LFWF_s}} and {\ref{tab_LFWF_v}} are derived from Ref. \cite{Maji:2016yqo}. These parameters are tabulated for both $u$ and $d$ quarks in Table \ref{tab_NC}.
	\begin{table}[h]
		\centering 
		\begin{tabular}{ |p{1.2cm}|p{1.9cm}|p{1.9cm}|p{1.9cm}|  }
			\hline
			~~$~~\nu$~~&~~$N_{S}$~~&~~$N_0^{\nu}$~~&~~$N_1^{\nu}$~~  \\
			\hline
			~~$~~u$~~&~~$2.0191$~~&~~$3.2050$~~&~~$0.9895$~~  \\
			~~$~~d$~~&~~$0$~~&~~$5.9423$~~&~~$1.1616$~~    \\
			\hline
		\end{tabular}
		\caption{Values of normalization constants $N_{i}^{2}$ which appears in Table {\ref{tab_LFWF_s}} and {\ref{tab_LFWF_v}}, corresponding to both up and down quarks.}
		\label{tab_NC} 
	\end{table}
	The parameter $\kappa$ for the AdS/QCD scale, occuring in Eq. (\ref{LFWF_phi}), has been assigned the value of $0.4~\mathrm{GeV}$ \cite{majiref28}. Following Ref. \cite{Chakrabarti:2019wjx}, the constituent quark mass ($m$) and proton mass ($M$) are taken to be $0.055~\mathrm{GeV}$ and $0.938~\mathrm{GeV}$ respectively.
	
	\section{Quark Correlator and TMDs}\label{sectmdsp}
	In the light-front formalism for SIDIS, the un-integrated quark-quark correlator can be defined as \cite{Maji:2017bcz}
	\be
	\Phi^{\nu [\Gamma]}(x,\textbf{p}_\perp;S)&=&\frac{1}{2}\int \frac{dz^- d^2z_T}{2(2\pi)^3} e^{ip.z} \langle P; S|\bar{\psi}^\nu (0)\Gamma \mathcal{W}_{[0,z]} \psi^\nu (z) |P;S\rangle\Bigg|_{z^+=0}, \label{TMDcor}
	\ee
	at equal light-front time $z^+=0$. Here the quark's longitudinal momentum fraction is represented by $x=p^+/P^+$. Proton's spin is $S$ and its momentum is denoted by $P$.

	We choose light-cone gauge $A^+=0$ and the frame is taken where the transverse momentum of the proton is $ P\equiv (P^+,\frac{M^2}{P^+},\textbf{0}_\perp)$, 
	and the momentum of the virtual photon is $q\equiv (x_B P^+, \frac{Q^2}{x_BP^+},\textbf{0})$, where $x_B= \frac{Q^2}{2P.q}$ is the Bjorken variable and $Q^2 = -q^2$. If the proton's helicity is $\lambda$, its spin components are denoted as $S^+ = \lambda \frac{P^+}{M},~ S^- = \lambda\frac{P^-}{M},$ and $ S_\perp $. In our work, the value of Wilson line is taken to be $1$. There are a total of $16$ TMDs in sub-leading twist case, $8$ of which are T-even and $8$ of which are T-odd. The sub-leading twist quark TMDs are projected in the form of Eq. (\ref{TMDcor}) according to Ref. \cite{Goeke:2005hb} and are represented as follows \big(color online: {\color{blue}blue: T-even}, {\color{red}red: T-odd}\big)
	\be
	\Phi^{\nu[1]}&=&\frac{M}{P^{+}}\left[{\color{blue}e^{\nu}\left(x, \textbf{p}_{\perp}^2\right)}-\frac{\varepsilon_{T}^{i j} p_{\perp i} S_{T j}}{M} {\color{red}e_{T}^{\perp\nu}\left(x, \textbf{p}_{\perp}^2\right)}\right], \label{Eqtmd1}\\
	\Phi^{\nu [i \gamma_{5}]}&=& \frac{M}{P^{+}}\left[\lambda~ {\color{red}e_{L}^{\nu}\left(x, \textbf{p}_{\perp}^2\right)}+\frac{\vec{\textbf{p}_{\perp}} \cdot \vec{S}_{T}}{M} {\color{red}e_{T}^{\nu}\left(x, \textbf{p}_{\perp}^2\right)}\right], \label{Eqtmd2}\\
	\Phi^{\nu [\gamma^{i}]}&=& \frac{M}{P^{+}}\bigg[\frac{{p}_{\perp}^{i}}{M}\left({\color{blue}f^{\perp \nu}\left(x, \textbf{p}_{\perp}^2\right)}-\frac{\varepsilon_{T}^{j k} {p}_{\perp j} S_{T k}}{M} {\color{red}f_{T}^{\perp ' \nu}\left(x, \textbf{p}_{\perp}^2\right)}\right)\bigg] \nonumber\\
	&&\left.+\frac{\varepsilon_{T}^{i j} {p}_{\perp j}}{M}\left(\lambda~{\color{red}f_{L}^{\perp\nu}\left(x, \textbf{p}_{\perp}^2\right)}+\frac{\vec{\textbf{p}_{\perp}} \cdot \vec{S}_{T}}{M} {\color{red}f_{T}^{\perp \nu}\left(x, \textbf{p}_{\perp}^2\right)}\right)\right],\label{Eqtmd3} \\
	\Phi^{\nu [\gamma^{i} \gamma_{5}]}&=& \frac{M}{P^{+}}\left[S_{T}^{i}~{\color{blue}g_{T}^{\prime\nu}\left(x, \textbf{p}_{\perp}^2\right)}+\frac{{p}_{\perp}^{i}}{M}\left(\lambda~{\color{blue}g_{L}^{\perp\nu}\left(x, \textbf{p}_{\perp}^2\right)}+\frac{\vec{\textbf{p}_{\perp}} \cdot \vec{S}_{T}}{M} {\color{blue}g_{T}^{\perp\nu}\left(x, \textbf{p}_{\perp}^2\right)}\right.\right) \nonumber\\
	&&-\left.\frac{\varepsilon_{T}^{i j} {p}_{\perp j}}{M} {\color{red}g^{\perp\nu}\left(x, \textbf{p}_{\perp}^2\right)}\right], \label{Eqtmd4}\\
	\Phi^{\nu [i \sigma^{i j} \gamma_{5}]}&=& \frac{M}{P^{+}}\left[\frac{S_{T}^{i} {p}_{\perp}^{j}-{p}_{\perp}^{i} S_{T}^{j}}{M} {\color{blue}h_{T}^{\perp\nu}\left(x, \textbf{p}_{\perp}^2\right)}-\varepsilon_{T}^{i j}~{\color{red}h^{\nu}\left(x, \textbf{p}_{\perp}^2\right)}\right], \label{Eqtmd5}\\
	\Phi^{\nu [i \sigma^{+-} \gamma_{5}]} &=& \frac{M}{P^{+}}{\left[\lambda~{\color{blue}h_{L}^{\nu}\left(x, \textbf{p}_{\perp}^2\right)}+\frac{\vec{\textbf{p}_{\perp}} \cdot \vec{S}_{T}}{M} {\color{blue}h_{T}^{\nu}\left(x, \textbf{p}_{\perp}^2\right)}\right] }. \label{Eqtmd6}
	\ee
	\par
	This study focuses on the sub-leading twist T-even TMDs. We have utilized the conventional notation for $\sigma^{i j}=i\left[\gamma^{i}, \gamma^{j}\right] / 2$, where $p_\perp^2=|\vec{p}_\perp|^2$. We have utilized the definition $\varepsilon_{\perp}^{i j}=\varepsilon^{-+i j}$, where $\varepsilon_\perp^{12}=-\varepsilon_\perp^{21}=1$ and is zero when $i$ and $j$ are identical. The indices $i$ and $j$ are used to indicate transverse directions.
	\section{Results}\label{secresults}
	\subsection{Overlap Form}\label{subsoverlap}
	It is a standard procedure to write the TMDs as wave functions that indicate the initial and final spin states of quarks and protons, a form known as the overlap form. To derive the overlap form of TMDs for the scalar diquark, we must substitute Eq. \eqref{fockSD}, with the appropriate polarization, in Eq. \eqref{TMDcor} via Eq. \eqref{PS_state}. After selecting a specific correlation (for $\Gamma= 1,~i \gamma_{5},~\gamma^{i},~\gamma^{i} \gamma_{5},~i \sigma^{i j} \gamma_{5}$ and $i \sigma^{+-} \gamma_{5}$) from Eqs. {\eqref{Eqtmd1}-\eqref{Eqtmd6}}, we may compute a specific TMD by selecting the proper combination of proton polarization. To calculate the longitudinally polarized TMD $h_{L}^{\nu}\left(x, \textbf{p}_{\perp}^2\right)$ in Eq. \eqref{Eqtmd6}, for example, we must use $\Gamma=i \sigma^{+-} \gamma_{5}$ and assume that the proton is longitudinally polarized in both the initial and final states. Thus, the sub-leading twist T-even TMDs for the scalar diquark may be written in terms of LFWFs as
	\begin{eqnarray}
		x \ e^{\nu(S)}(x,\textbf{p}_{\perp}^2)&=&\frac{1}{16\pi^3}\frac{m}{M}\bigg[|\psi ^{+\nu}_+(x,\textbf{p}_{\perp})|^2+|\psi ^{ + \nu}_-(x,\textbf{p}_{\perp})|^2\bigg],\label{es}  \\
		{x}~f^{\perp\nu(S)}(x,\textbf{p}_{\perp}^2)&=&\frac{1}{16\pi^3}\bigg[|\psi ^{+\nu}_+(x,\textbf{p}_{\perp})|^2+|\psi ^{ + \nu}_-(x,\textbf{p}_{\perp})|^2\bigg], \\
		x \ p_x g_{L}^{\perp\nu(S)}(x, {\bf p_\perp^2})&=&\frac{1}{32\pi^3}\bigg(p_x\bigg[|\psi ^{+\nu}_+(x,\textbf{p}_{\perp})|^2-|\psi ^{ + \nu}_-(x,\textbf{p}_{\perp})|^2\bigg] \nonumber\\
		&&+ m \bigg[\psi ^{+\nu \dagger}_+(x,\textbf{p}_{\perp}) \psi ^{+\nu}_-(x,\textbf{p}_{\perp}) + 
		\psi ^{+\nu \dagger}_-(x,\textbf{p}_{\perp}) \psi ^{+\nu}_+(x,\textbf{p}_{\perp})\bigg]\bigg),\\ 
		{x M}\bigg(g_{T}^{'\nu(S)}(x, {\bf p_\perp^2})+\frac{\textbf{p}_{x}^2}{M^2} g_{T}^{\perp\nu(S)}(x, {\bf p_\perp^2})\bigg)&=&  \frac{1}{32\pi^3}\bigg(\textbf{p}_{x}\bigg[\psi ^{+\nu \dagger}_+(x,\textbf{p}_{\perp})\psi^{- \nu}_+(x,\textbf{p}_{\perp}) - \psi ^{+\nu \dagger}_-(x,\textbf{p}_{\perp}) \psi ^{-\nu }_-(x,\textbf{p}_{\perp}) \nonumber\\
		&&+ \psi ^{-\nu \dagger}_+(x,\textbf{p}_{\perp}) \psi ^{+\nu}_+(x,\textbf{p}_{\perp})-\psi^{- \dagger \nu}_-(x,\textbf{p}_{\perp}) \psi ^{+\nu}_-(x,\textbf{p}_{\perp})\bigg]\nonumber\\
		&& +m\bigg[\psi ^{+\nu \dagger}_+(x,\textbf{p}_{\perp})\psi^{- \nu}_-(x,\textbf{p}_{\perp}) + \psi ^{+\nu \dagger}_-(x,\textbf{p}_{\perp})\psi^{- \nu}_+(x,\textbf{p}_{\perp}) \nonumber \\
		&&+ \psi ^{-\nu \dagger}_+(x,\textbf{p}_{\perp})\psi^{+ \nu}_-(x,\textbf{p}_{\perp})+ \psi ^{-\nu \dagger}_-(x,\textbf{p}_{\perp})\psi^{+ \nu}_+(x,\textbf{p}_{\perp})\bigg] \bigg),\\
		{x} \frac{{\bf p_x} \ {\bf p_y}}{M} g_{T}^{\perp(S)}(x, {\bf p_\perp^2})&=& \frac{1}{32\pi^3}  \bigg({\bf p_y}\bigg[\psi ^{+\nu \dagger}_+(x,\textbf{p}_{\perp})\psi^{- \nu}_+(x,\textbf{p}_{\perp}) - \psi ^{+\nu \dagger}_-(x,\textbf{p}_{\perp}) \psi ^{-\nu}_-(x,\textbf{p}_{\perp}) \nonumber\\
		&& + \psi ^{-\nu \dagger}_+(x,\textbf{p}_{\perp}) \psi ^{+\nu}_+(x,\textbf{p}_{\perp})-\psi^{- \dagger \nu}_-(x,\textbf{p}_{\perp}) \psi ^{+\nu}_-(x,\textbf{p}_{\perp})\bigg] \nonumber\\
		&& + \iota m \bigg[
		\psi ^{+\nu \dagger}_+(x,\textbf{p}_{\perp}) \psi ^{-\nu }_-(x,\textbf{p}_{\perp})- \psi ^{-\nu \dagger}_-(x,\textbf{p}_{\perp}) \psi ^{+\nu}_+(x,\textbf{p}_{\perp})\nonumber\\
		&& +\psi ^{+\nu \dagger}_-(x,\textbf{p}_{\perp}) \psi ^{-\nu }_+(x,\textbf{p}_{\perp})- \psi ^{-\nu \dagger}_+(x,\textbf{p}_{\perp}) \psi ^{+\nu}_-(x,\textbf{p}_{\perp})\bigg]\bigg), \\
		{x M} h_{L}^{\nu(S)}(x, {\bf p_\perp^2}) &=&\frac{1}{16\pi^3} \bigg(m\bigg[|\psi ^{+\nu}_+(x,\textbf{p}_{\perp})|^2-|\psi ^{ + \nu}_-(x,\textbf{p}_{\perp})|^2\bigg] \nonumber\\
		&& - ({\textbf{p}_{x}}-\iota {\textbf{p}_{y}})\bigg[\psi ^{+\nu \dagger}_+(x,\textbf{p}_{\perp})\psi^{+ \nu}_-(x,\textbf{p}_{\perp})\bigg] \nonumber \\
		&&- ({\textbf{p}_{x}}+\iota {\textbf{p}_{y}}) \bigg[\psi ^{+\nu \dagger}_-(x,\textbf{p}_{\perp})\psi^{+ \nu}_+(x,\textbf{p}_{\perp})\bigg]\bigg), \\
		{x}~{\textbf{p}_{x}}~h_{T}^{\nu(S)}(x, {\bf p_\perp^2}) &=&\frac{1}{16\pi^3}\bigg(m\bigg[\psi ^{+\nu \dagger}_+(x,\textbf{p}_{\perp})\psi^{- \nu}_+(x,\textbf{p}_{\perp}) - \psi ^{+\nu \dagger}_-(x,\textbf{p}_{\perp}) \psi ^{-\nu }_-(x,\textbf{p}_{\perp}) \nonumber\\
		&&+ \psi ^{-\nu \dagger}_+(x,\textbf{p}_{\perp}) \psi ^{+\nu}_+(x,\textbf{p}_{\perp})-\psi^{- \dagger \nu}_-(x,\textbf{p}_{\perp}) \psi ^{+\nu}_-(x,\textbf{p}_{\perp})\bigg]\nonumber \\
		&& -({\textbf{p}_{x}}-\iota {\textbf{p}_{y}})\bigg[\psi ^{+\nu \dagger}_+(x,\textbf{p}_{\perp})\psi^{- \nu}_-(x,\textbf{p}_{\perp}) + \psi ^{-\nu \dagger}_+(x,\textbf{p}_{\perp})\psi^{+ \nu}_-(x,\textbf{p}_{\perp})\bigg] \nonumber \\
		&& -({\textbf{p}_{x}}+\iota {\textbf{p}_{y}})\bigg[\psi ^{+\nu \dagger}_-(x,\textbf{p}_{\perp})\psi^{- \nu}_+(x,\textbf{p}_{\perp}) + \psi ^{-\nu \dagger}_-(x,\textbf{p}_{\perp})\psi^{+ \nu}_+(x,\textbf{p}_{\perp})\bigg]\bigg),\label{ht2s} 
	\nonumber \\
		x \ {\textbf{p}_{y}} \ h_{T}^{\perp\nu(S)}(x, {\bf p_\perp^2}) &=& \frac{1}{32\pi^3} \bigg(({\iota}{\textbf{p}_{x}}+ {\textbf{p}_{y}})\bigg[\psi ^{+\nu \dagger}_+(x,\textbf{p}_{\perp})\psi^{- \nu}_-(x,\textbf{p}_{\perp}) + \psi ^{-\nu \dagger}_+(x,\textbf{p}_{\perp})\psi^{+ \nu}_-(x,\textbf{p}_{\perp})\bigg] \nonumber \\
		&& -({\iota}{\textbf{p}_{x}}-{\textbf{p}_{y}})\bigg[\psi ^{+\nu \dagger}_-(x,\textbf{p}_{\perp})\psi^{- \nu}_+(x,\textbf{p}_{\perp}) + \psi ^{-\nu \dagger}_-(x,\textbf{p}_{\perp})\psi^{+ \nu}_+(x,\textbf{p}_{\perp})\bigg]\bigg).
		\label{htperp2s}\nonumber\\
	\end{eqnarray}
	Now, to get the overlap form of TMDs for the vector diquark, we  substitute Eq. \eqref{fockVD}, with suitable polarization, in Eq. \eqref{TMDcor} via Eq. \eqref{PS_state}. The TMDs in terms of LFWFs for the vector diquark can be written  as
	\begin{eqnarray}
		{x}~e^{\nu(A)}(x,\textbf{p}_{\perp}^2) &=&\sum_{\lambda_D} \frac{1}{16\pi^3}\frac{m}{M}\bigg[|\psi ^{+\nu}_{+\lambda_D}(x,\textbf{p}_{\perp})|^2+|\psi ^{ + \nu}_{-\lambda_D}(x,\textbf{p}_{\perp})|^2\bigg], \label{ev}\\
		{x}~f^{\perp\nu(A)}(x,\textbf{p}_{\perp}^2)&=&\sum_{\lambda_D} \frac{1}{16\pi^3}\bigg[|\psi ^{+\nu}_{+\lambda_D}(x,\textbf{p}_{\perp})|^2+|\psi ^{ + \nu}_{-\lambda_D}(x,\textbf{p}_{\perp})|^2\bigg],\label{fperpv} \\
		{x}~{\textbf{p}_{x}} g_{L}^{\perp\nu(A)}(x, {\bf p_\perp^2}) &=& \sum_{\lambda_D}\frac{1}{32\pi^3}\bigg({\textbf{p}_{x}}\bigg[|\psi ^{+\nu}_{+\lambda_D}(x,\textbf{p}_{\perp})|^2-|\psi ^{ + \nu}_{-\lambda_D}(x,\textbf{p}_{\perp})|^2\bigg] \nonumber\\
		&& + m\bigg[\psi ^{+\nu \dagger}_{+\lambda_D}(x,\textbf{p}_{\perp})\psi^{+ \nu}_{-\lambda_D}(x,\textbf{p}_{\perp}) + \psi ^{+\nu \dagger}_{-\lambda_D}(x,\textbf{p}_{\perp})\psi^{+ \nu}_{+\lambda_D}(x,\textbf{p}_{\perp})\bigg]\bigg), \nonumber\\
		\label{glperp1v} \\
		{x M}\bigg(g_{T}^{'\nu(A)}(x, {\bf p_\perp^2})+\frac{\textbf{p}_{x}^2}{M^2}g_{T}^{\perp (A)}(x, {\bf p_\perp^2})\bigg) &=&\frac{1}{32\pi^3}\bigg(\textbf{p}_{x}\bigg[\psi ^{+\nu \dagger}_{+0}(x,\textbf{p}_{\perp})\psi^{- \nu}_{+0}(x,\textbf{p}_{\perp}) - \psi ^{+\nu \dagger}_{-0}(x,\textbf{p}_{\perp})\psi^{- \nu}_{-0}(x,\textbf{p}_{\perp})\nonumber \\
		&&+ \psi ^{-\nu \dagger}_{+0}(x,\textbf{p}_{\perp})\psi^{+ \nu}_{+0}(x,\textbf{p}_{\perp})- \psi ^{-\nu \dagger}_{-0}(x,\textbf{p}_{\perp})\psi^{+ \nu}_{-0}(x,\textbf{p}_{\perp})\bigg]  \nonumber \\
		&& +m\bigg[\psi ^{+\nu \dagger}_{+0}(x,\textbf{p}_{\perp})\psi^{- \nu}_{-0}(x,\textbf{p}_{\perp}) + \psi ^{+\nu \dagger}_{-0}(x,\textbf{p}_{\perp})\psi^{- \nu}_{+0}(x,\textbf{p}_{\perp})\nonumber \\
		&&+\psi ^{-\nu \dagger}_{+0}(x,\textbf{p}_{\perp})\psi^{+ \nu}_{-0}(x,\textbf{p}_{\perp}) + \psi ^{-\nu \dagger}_{-0}(x,\textbf{p}_{\perp})\psi^{+ \nu}_{+0}(x,\textbf{p}_{\perp}) \bigg] \bigg), \label{gt1v}  \\
		{x}~\frac{{\bf p_x} \ {\bf p_y}}{M}~g_{T}^{\perp\nu (A)}(x, {\bf p_\perp^2}) &=& \frac{1}{32\pi^3}\bigg[\textbf{p}_{y}\bigg[\psi ^{+\nu \dagger}_{+0}(x,\textbf{p}_{\perp})\psi^{- \nu}_{+0}(x,\textbf{p}_{\perp}) - \psi ^{+\nu \dagger}_{-0}(x,\textbf{p}_{\perp})\psi^{- \nu}_{-0}(x,\textbf{p}_{\perp})\nonumber \\
		&+& \psi^{-\nu \dagger}_{+0}(x,\textbf{p}_{\perp})\psi^{+ \nu}_{+0}(x,\textbf{p}_{\perp})- \psi^{-\nu \dagger}_{-0}(x,\textbf{p}_{\perp})\psi^{+ \nu}_{-0}(x,\textbf{p}_{\perp})\bigg]  \nonumber \\
		&& -\iota m\bigg[\psi ^{+\nu \dagger}_{+0}(x,\textbf{p}_{\perp})\psi^{- \nu}_{-0}(x,\textbf{p}_{\perp}) - \psi ^{+\nu \dagger}_{-0}(x,\textbf{p}_{\perp})\psi^{- \nu}_{+0}(x,\textbf{p}_{\perp}) \nonumber \\
		&&+\psi ^{-\nu \dagger}_{+0}(x,\textbf{p}_{\perp})\psi^{+ \nu}_{-0}(x,\textbf{p}_{\perp}) -\psi ^{-\nu \dagger}_{-0}(x,\textbf{p}_{\perp})\psi^{+ \nu}_{+0}(x,\textbf{p}_{\perp}) \bigg] \bigg],  \label{gtperp2v}
		\\
		{x M}~h_{L}^{\nu(A)}(x, {\bf p_\perp^2}) &=&\sum_{\lambda_D}\frac{1}{16\pi^3}\bigg(m\bigg[|\psi ^{+\nu}_{+\lambda_D}(x,\textbf{p}_{\perp})|^2-|\psi ^{ + \nu}_{-\lambda_D}(x,\textbf{p}_{\perp})|^2\bigg] \nonumber \\
		&& - ({\textbf{p}_{x}}-\iota {\textbf{p}_{y}})\bigg[\psi ^{+\nu \dagger}_{+\lambda_D}(x,\textbf{p}_{\perp})\psi^{+ \nu}_{-\lambda_D}(x,\textbf{p}_{\perp})\bigg]- ({\textbf{p}_{x}}+\iota {\textbf{p}_{y}})\nonumber\\
		&& \bigg[\psi ^{+\nu \dagger}_{-\lambda_D}(x,\textbf{p}_{\perp})\psi^{+ \nu}_{+\lambda_D}(x,\textbf{p}_{\perp})\bigg]\bigg), \label{hlv} \\
		{x}~{\textbf{p}_{x}}~h_{T}^{\nu(A)}(x, {\bf p_\perp^2}) &=&\frac{1}{16\pi^3}\bigg(m\bigg[\psi ^{+\nu \dagger}_{+0}(x,\textbf{p}_{\perp})\psi^{- \nu}_{+0}(x,\textbf{p}_{\perp}) - \psi ^{+\nu \dagger}_{-0}(x,\textbf{p}_{\perp})\psi^{- \nu}_{-0}(x,\textbf{p}_{\perp})\nonumber \\
		&& +\psi ^{-\nu \dagger}_{+0}(x,\textbf{p}_{\perp})\psi^{+ \nu}_{+0}(x,\textbf{p}_{\perp}) -\psi ^{-\nu \dagger}_{-0}(x,\textbf{p}_{\perp})\psi^{+ \nu}_{-0}(x,\textbf{p}_{\perp}) \bigg] \nonumber\\
		&& -({\textbf{p}_{x}}-\iota {\textbf{p}_{y}})\bigg[\psi ^{+\nu \dagger}_{+0}(x,\textbf{p}_{\perp})\psi^{- \nu}_{-0}(x,\textbf{p}_{\perp}) + \psi ^{-\nu \dagger}_{+0}(x,\textbf{p}_{\perp})\psi^{+ \nu}_{-0}(x,\textbf{p}_{\perp})\bigg] \nonumber \\
		&& -({\textbf{p}_{x}}+\iota {\textbf{p}_{y}})\bigg[\psi ^{+\nu \dagger}_{-0}(x,\textbf{p}_{\perp})\psi^{- \nu}_{+0}(x,\textbf{p}_{\perp}) + \psi ^{-\nu \dagger}_{-0}(x,\textbf{p}_{\perp})\psi^{+ \nu}_{+0}(x,\textbf{p}_{\perp})\bigg]\bigg),
		\nonumber\\
		\label{ht1v} 
			\end{eqnarray}
	\begin{eqnarray}
		{x}~{\textbf{p}_{y}}~h_{T}^{\perp\nu(A)}(x, {\bf p_\perp^2}) &=&\frac{1}{32\pi^3} \bigg(({\iota}{\textbf{p}_{x}}+ {\textbf{p}_{y}})\bigg[\psi ^{+\nu \dagger}_{+0}(x,\textbf{p}_{\perp})\psi^{- \nu}_{-0}(x,\textbf{p}_{\perp}) + \psi ^{-\nu \dagger}_{+0}(x,\textbf{p}_{\perp})\psi^{+ \nu}_{-0}(x,\textbf{p}_{\perp})\bigg] \nonumber\\
		&&  -({\iota}{\textbf{p}_{x}}-{\textbf{p}_{y}})\bigg[\psi ^{+\nu \dagger}_{-0}(x,\textbf{p}_{\perp})\psi^{- \nu}_{+0}(x,\textbf{p}_{\perp}) + \psi ^{-\nu \dagger}_{-0}(x,\textbf{p}_{\perp})\psi^{+ \nu}_{+0}(x,\textbf{p}_{\perp})\bigg]\bigg),
		\nonumber\\
		\label{htperp1v}
	\end{eqnarray}
	where the summation over helicity of the vector diquark is taken, $\lambda_D= 0,\pm 1$.
	\subsection{Explicit Expressions of TMDs}\label{subsexplicit}
	We have derived the explicit formulations for sub-leading twist T-even TMDs by inserting the values of the LFWFs for the scalar and vector diquarks from Tables \ref{tab_LFWF_s} and \ref{tab_LFWF_v} into Eqs. {\eqref{es}-\eqref{htperp1v}} as
	\begin{eqnarray}
		x e^{\nu}(x, {\bf p_\perp^2}) &=& \frac{1}{16\pi^3}\Bigg({C_{S}^{2} N_s^2}+{C_{A}^{2}}  \bigg(\frac{2}{3}|N_1^\nu|^2 + \frac{1}{3} |N_0^\nu|^2\bigg)\Bigg) \frac{m}{M} \bigg[|\varphi_1^\nu|^2 + \frac{p_\perp^2}{x^2 M^2}|\varphi_2^\nu|^2\bigg], \label{ef} \\
		x f^{\perp\nu}(x, {\bf p_\perp^2}) &=&  \frac{1}{16 \pi^3} \Bigg({C_{S}^{2} N_s^2} +{C_{A}^{2}} \bigg(\frac{2}{3}|N_1^\nu|^2 + \frac{1}{3} |N_0^\nu|^2\bigg)\Bigg) \bigg[|\varphi_1^\nu|^2 + \frac{p_\perp^2}{x^2 M^2}|\varphi_2^\nu|^2\bigg], \label{fperpf}\\
		x g_{L}^{\perp\nu}(x, {\bf p_\perp^2}) &=&  \frac{1}{32 \pi^3}\Bigg({C_{S}^{2} N_s^2}+{C_{A}^{2}} \bigg(-\frac{2}{3}|N_1^\nu|^2 + \frac{1}{3} |N_0^\nu|^2\bigg)\Bigg) \bigg[|\varphi_1^\nu|^2 - \frac{p_\perp^2}{x^2 M^2}|\varphi_2^\nu|^2-\frac{2 m}{x M}|\varphi_1^\nu||\varphi_2^\nu| \bigg], \label{glperpf}\\
		x g_{T}^{'\nu}(x, {\bf p_\perp^2}) &=&  \frac{1}{16 \pi^3}\Bigg( {C_{S}^{2} N_s^2}-\frac{1}{3} {{C_{A}^{2}} |N_0^\nu|^2}\Bigg)\frac{m}{M} \bigg[|\varphi_1^\nu|^2 + \frac{p_\perp^2}{x^2 M^2}|\varphi_2^\nu|^2\bigg], \label{gtf}\\
		x g_{T}^{\perp\nu}(x, {\bf p_\perp^2}) &=&  \frac{1}{8 \pi^3} \Bigg({C_{S}^{2} N_s^2}-\frac{1}{3} {{C_{A}^{2}} |N_0^\nu|^2})\Bigg)\bigg[ \frac{1}{x}|\varphi_1^\nu||\varphi_2^\nu|- \frac{m}{x^2 M}|\varphi_2^\nu|^2 \bigg],\label{gtperpf}\\
		x h_{L}^{\nu}(x, {\bf p_\perp^2}) &=& \frac{1}{16 \pi^3} \Bigg( {C_{S}^{2} N_s^2}+ {C_{A}^{2}}\bigg(-\frac{2}{3}|N_1^\nu|^2 + \frac{1}{3} |N_0^\nu|^2\bigg) \Bigg)   \frac{1}{M} \bigg[m\bigg(|\varphi_1^\nu|^2 - \frac{p_\perp^2}{x^2 M^2}|\varphi_2^\nu|^2\bigg) \nonumber \\
		&&+\frac{2 p_\perp^2}{x M}|\varphi_1^\nu||\varphi_2^\nu| \bigg], \label{hlf} \\
		x h_{T}^{\nu}(x, {\bf p_\perp^2}) &=&  \frac{1}{8 \pi^3} \Bigg( -{C_{S}^{2} N_s^2}+\frac{1}{3} {{C_{A}^{2}} |N_0^\nu|^2} \Bigg) \bigg[|\varphi_1^\nu|^2 - \frac{p_\perp^2}{x^2 M^2}|\varphi_2^\nu|^2-\frac{2 m}{x M}|\varphi_1^\nu||\varphi_2^\nu| \bigg],  \label{htf}\\
		x h_{T}^{\perp\nu}(x, {\bf p_\perp^2}) &=&  \frac{1}{16 \pi^3}
		\Bigg( {C_{S}^{2} N_s^2}-\frac{1}{3} {{C_{A}^{2}} |N_0^\nu|^2}\Bigg) \bigg[|\varphi_1^\nu|^2 + \frac{p_\perp^2}{x^2 M^2}|\varphi_2^\nu|^2\bigg]. \label{htperpf}
	\end{eqnarray}
	\begin{figure*}
		\centering
		\begin{minipage}[c]{0.98\textwidth}
			(a)\includegraphics[width=7.5cm]{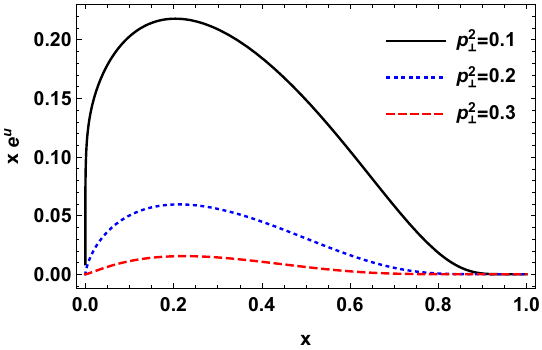}
			\hspace{0.05cm}
			(b)\includegraphics[width=7.5cm]{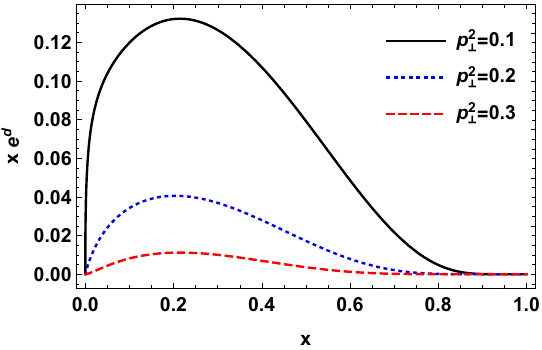}
			\hspace{0.05cm}
			(c)\includegraphics[width=7.5cm]{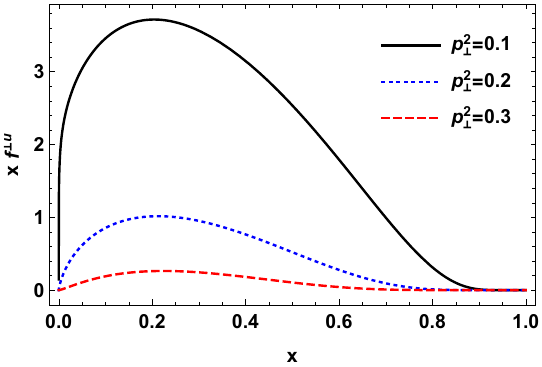}
			\hspace{0.05cm}
			(d)\includegraphics[width=7.5cm]{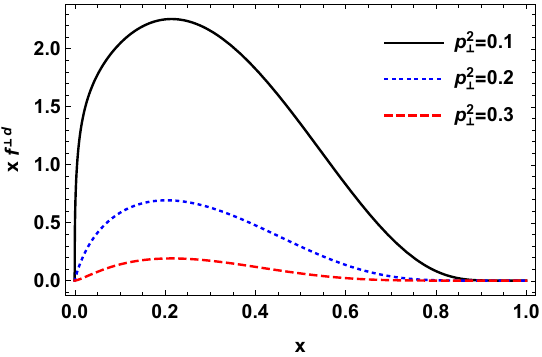}
			\hspace{0.05cm}\\
		\end{minipage}
		\caption{\label{figTMDvx1} (Color online) The unpolarized TMDs  $x e^{\nu}(x, {\bf p_\perp^2})$ and $x f^{\perp\nu}(x, {\bf p_\perp^2})$  plotted with respect to $x$ at different values of $ {\bf p_\perp^2}$. The left and right column correspond to $u$ and $d$ quarks respectively.}
	\end{figure*}
	\section{Discussion}\label{secdiscussion}
\begin{figure*}
		\centering
		\begin{minipage}[c]{0.98\textwidth}
			(a)\includegraphics[width=7.5cm]{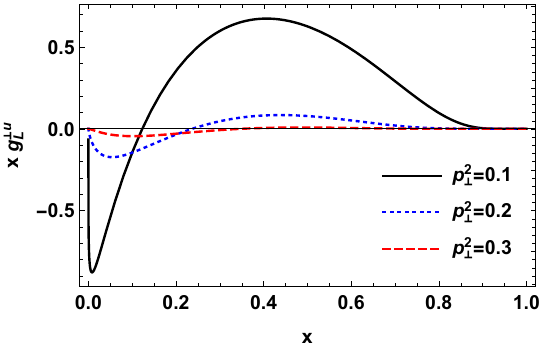}
			\hspace{0.05cm}
			(b)\includegraphics[width=7.5cm]{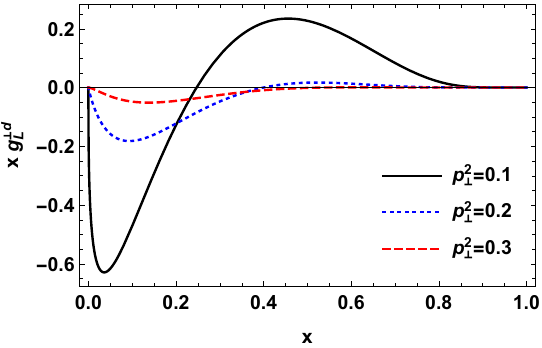}
			\hspace{0.05cm}
			(c)\includegraphics[width=7.5cm]{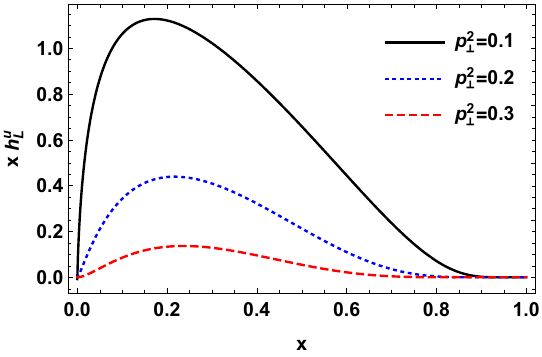}
			\hspace{0.05cm}
			(d)\includegraphics[width=7.5cm]{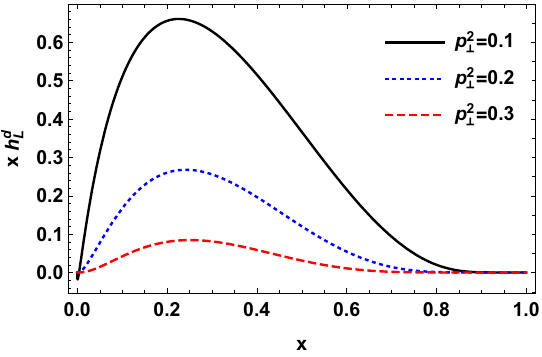}
			\hspace{0.05cm}\\
		\end{minipage}
		\caption{\label{figTMDvx2} (Color online) The longitudinally polarized TMDs  $x g_{L}^{\perp\nu}(x, {\bf p_\perp^2})$ and $~x h_{L}^{\nu}(x, {\bf p_\perp^2})$  plotted with respect to $x$ at different values of $ {\bf p_\perp^2}$. The left and right column correspond to $u$ and $d$ quarks respectively.}
	\end{figure*}

\begin{figure*}
		\centering
		\begin{minipage}[c]{0.98\textwidth}
			(a)\includegraphics[width=7.48cm]{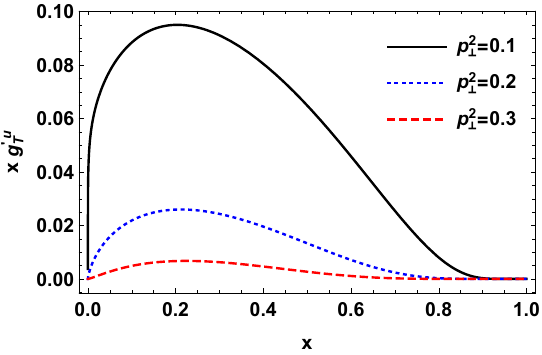}
			\hspace{0.05cm}
			(b)\includegraphics[width=7.48cm]{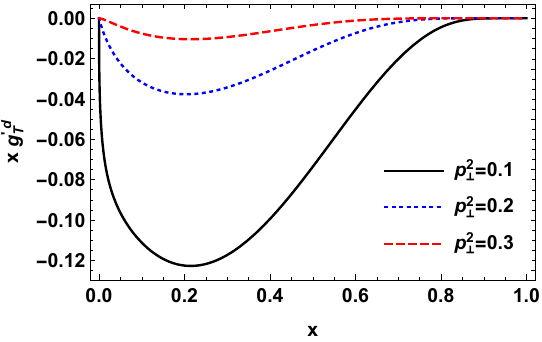}
			\hspace{0.05cm}
			(c)\includegraphics[width=7.5cm]{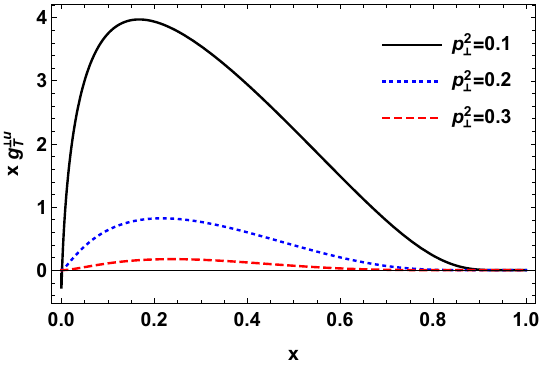}
			\hspace{0.05cm}
			(d)\includegraphics[width=7.5cm]{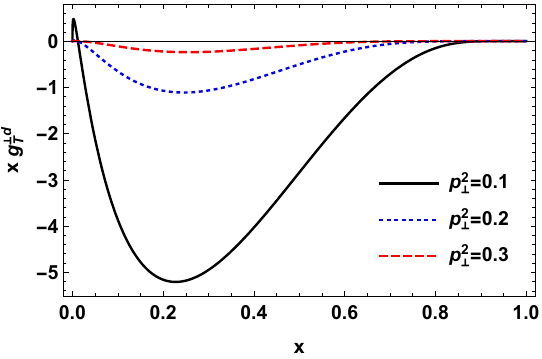}
			\hspace{0.05cm}
			(e)\includegraphics[width=7.5cm]{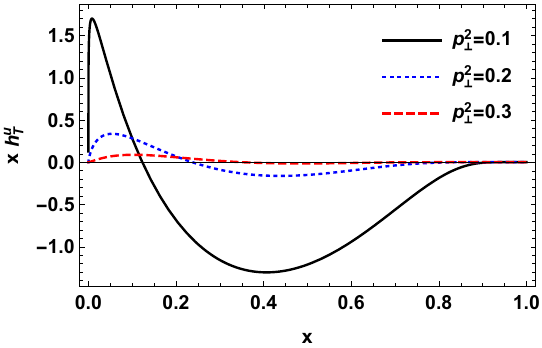}
			\hspace{0.05cm}
			(f)\includegraphics[width=7.5cm]{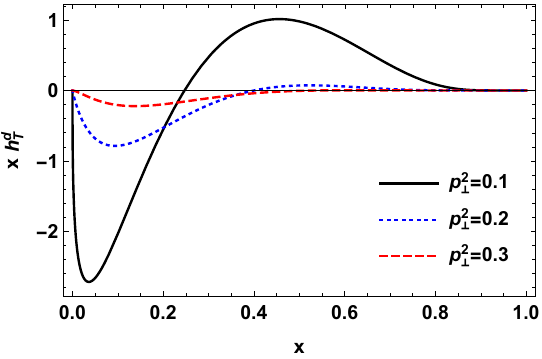}
			\hspace{0.05cm}
			(g)\includegraphics[width=7.48cm]{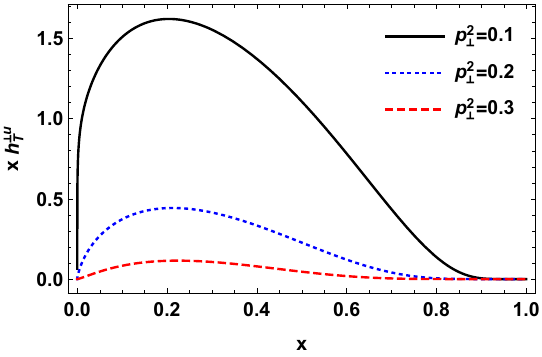}
			\hspace{0.05cm}
			(h)\includegraphics[width=7.48cm]{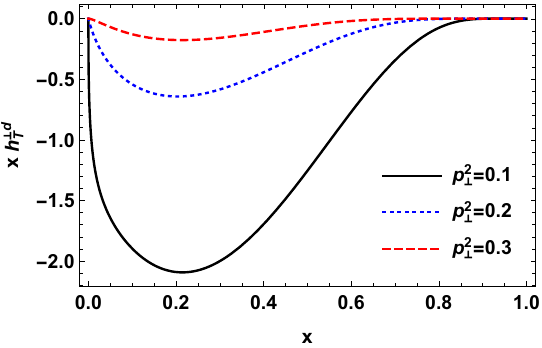}
			\hspace{0.05cm}\\
		\end{minipage}
		\caption{\label{figTMDvx3} (Color online) The transversely polarized TMDs  $x g_{T}^{'\nu}(x, {\bf p_\perp^2}), x g_{T}^{\perp\nu}(x, {\bf p_\perp^2}),~ x h_{T}^{\nu}(x, {\bf p_\perp^2})$ and $ x h_{T}^{\perp\nu}(x, {\bf p_\perp^2})$  plotted with respect to $x$ at different values of $ {\bf p_\perp^2}$. The left and right column correspond to $u$ and $d$ quarks respectively.}
	\end{figure*}

	To have a closer look on TMDs, we have plotted them with respect to $x$ at different values of $ {\bf p_\perp^2}$. In Figs. \ref{figTMDvx1}, \ref{figTMDvx2} and \ref{figTMDvx3}, TMDs for unpolarized $\bigg(x e^{\nu}(x, {\bf p_\perp^2})$ and $x f^{\perp\nu}(x, {\bf p_\perp^2})\bigg)$,
	longitudinally polarized $\bigg(x g_{L}^{\perp\nu}(x, {\bf p_\perp^2})$ and $~x h_{L}^{\nu}(x, {\bf p_\perp^2})\bigg)$
	and the transversely polarized $\bigg(x g_{T}^{'\nu}(x, {\bf p_\perp^2}), x g_{T}^{\perp\nu}(x, {\bf p_\perp^2}),$ $ x h_{T}^{\nu}(x, {\bf p_\perp^2})$ and $ x h_{T}^{\perp\nu}(x, {\bf p_\perp^2})\bigg)$ have been plotted with respect to $x$ at different discrete values of $ {\bf p_\perp^2}$, i.e., ${\bf p_\perp^2}=0.1 {\ \rm GeV}^2$ (black curve), ${\bf p_\perp^2}=0.2 {\ \rm GeV}^2$ (dotted blue curve) and ${\bf p_\perp^2}=0.3 {\ \rm GeV}^2$ (dashed red curve). The left column and the right column correspond, in order, to the $u$ and $d$ quark. Out of the eight T-even TMDs, only two TMDs i.e., $x e^{\nu}(x, {\bf p_\perp^2})$ and $ x f^{\perp\nu}(x, {\bf p_\perp^2})$ are unpolarized. Both of these sub-leading twist TMDs are related to the leading twist TMD $f_1^q(x, {\bf p_\perp})$ via Eqs. \eqref{ec} and \eqref{fpc}, considering the fact that the tilde terms are zero. With an increase in the longitudinal momentum fraction $x$, the TMDs $x e^{\nu}(x, {\bf p_\perp^2})$ and $ x f^{\perp\nu}(x, {\bf p_\perp^2})$ first increases and then decreases to follow a trend as shown in Fig. \ref{figTMDvx1} (a) to \ref{figTMDvx1} (d). The TMD $x e^{\nu}(x, {\bf p_\perp^2})$ is sizeable, which is quite explainable as our model operates at low $Q^2$. Similar type of behavior has been seen in the case of $ x f^{\perp\nu}(x, {\bf p_\perp^2})$.
	
	In Fig. \ref{figTMDvx2}, the results of longitudinally polarized TMDs $x g_{L}^{\perp\nu}(x, {\bf p_\perp^2})$ and $~x h_{L}^{\nu}(x, {\bf p_\perp^2})$ are shown. $x g_{L}^{\perp\nu}(x, {\bf p_\perp^2})$ shows a node around
	$x=0.12$ and $x=0.25$ for $u$ and $d$ quarks at $p_\perp^2=0.1$ respectively as shown in Fig. \ref{figTMDvx2} (a) and \ref{figTMDvx2} (b).
	Similar type of results have been seen in the CPM \cite{Bastami:2020rxn} and the bag model \cite{Avakian:2010br}. A result of $ g_{L}^{\perp\nu}(x, {\bf p_\perp^2})= -  h_{T}^{\nu}(x, {\bf p_\perp^2})$ has also been observed in CPM \cite{Bastami:2020rxn} but in our case such relation is not observed. However, from the plots of $ g_{L}^{\perp\nu}(x, {\bf p_\perp^2})$ and $ h_{T}^{\nu}(x, {\bf p_\perp^2})$, it is clear that they show an exactly opposite behavior to each other for $u$ quark but for $d$ quark they differ only by magnitude. This type of result indicates presence of some quark model symmetry. 
	\begin{figure*}
		\centering
		\begin{minipage}[c]{0.98\textwidth}
			(a)\includegraphics[width=7.5cm]{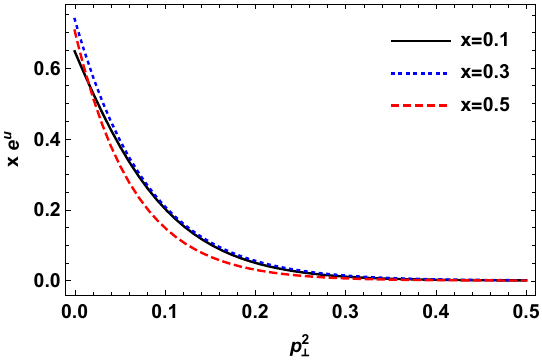}
			\hspace{0.05cm}
			(b)\includegraphics[width=7.5cm]{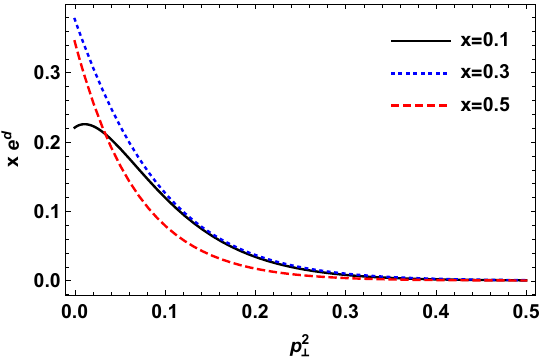}
			\hspace{0.05cm}
			(c)\includegraphics[width=7.5cm]{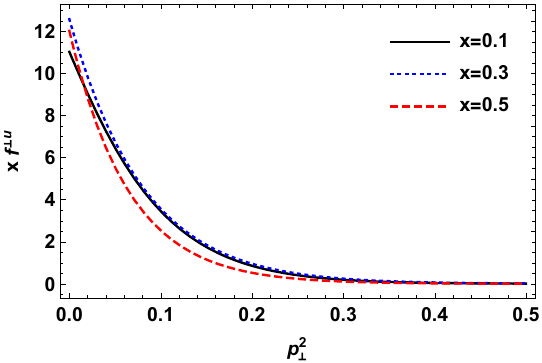}
			\hspace{0.05cm}
			(d)\includegraphics[width=7.5cm]{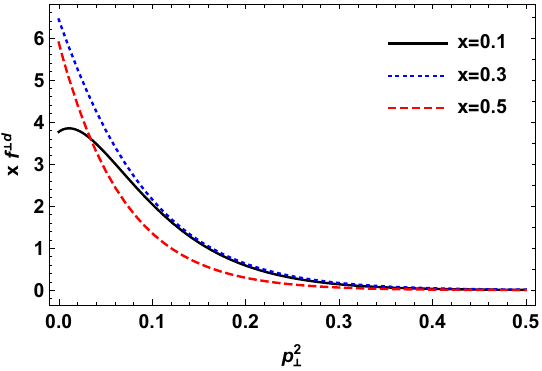}
			\hspace{0.05cm}\\
		\end{minipage}
		\caption{\label{figTMDvp1} (Color online) The unpolarized TMDs  $x e^{\nu}(x, {\bf p_\perp^2})$ and $x f^{\perp\nu}(x, {\bf p_\perp^2})$ plotted with respect to ${\bf p_\perp^2}$ at different values of $x$. The left and right column correspond to $u$ and $d$ quarks respectively.}
	\end{figure*}
	\begin{figure*}
		\centering
		\begin{minipage}[c]{0.98\textwidth}
			(a)\includegraphics[width=7.5cm]{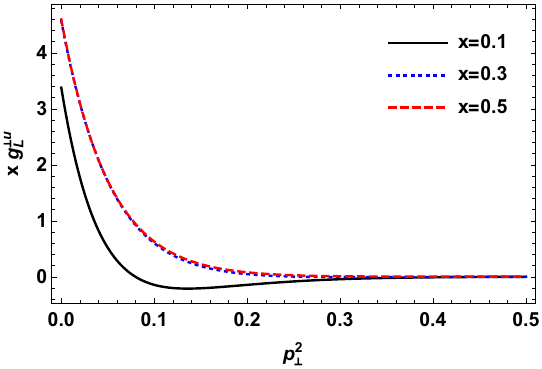}
			\hspace{0.05cm}
			(b)\includegraphics[width=7.5cm]{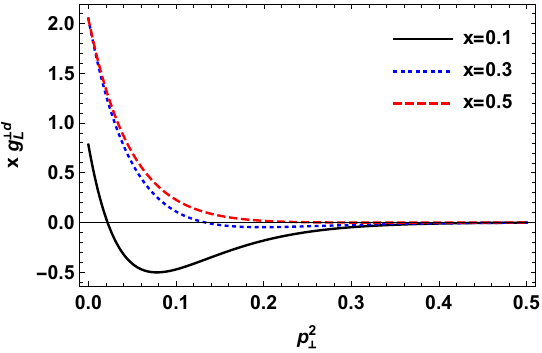}
			\hspace{0.05cm}
			(c)\includegraphics[width=7.5cm]{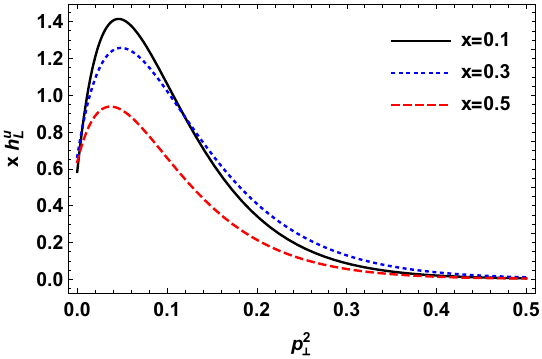}
			\hspace{0.05cm}
			(d)\includegraphics[width=7.5cm]{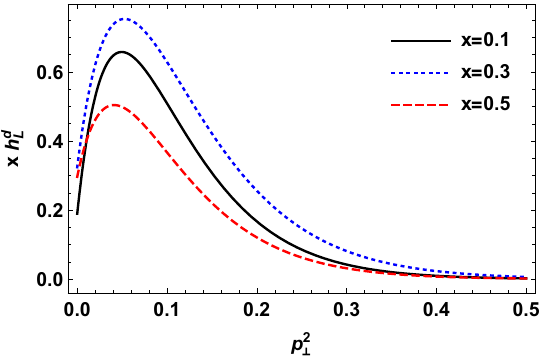}
			\hspace{0.05cm}\\
		\end{minipage}
		\caption{\label{figTMDvp2} (Color online) The longitudinally polarized TMDs $x g_{L}^{\perp\nu}(x, {\bf p_\perp^2})$ and $~x h_{L}^{\nu}(x, {\bf p_\perp^2})$ plotted with respect to ${\bf p_\perp^2}$ at different values of $x$. The left and right column correspond to $u$ and $d$ quarks respectively.}
	\end{figure*}
	\begin{figure*}
		\centering
		\begin{minipage}[c]{0.98\textwidth}
			(a)\includegraphics[width=7.48cm]{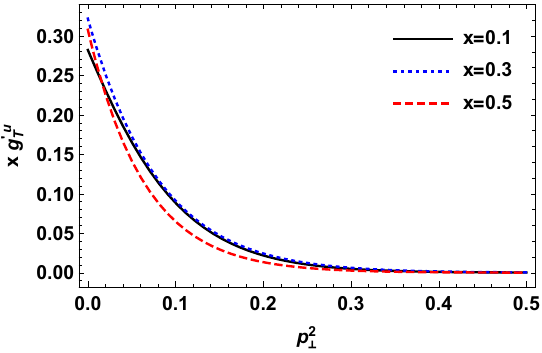}
			\hspace{0.05cm}
			(b)\includegraphics[width=7.48cm]{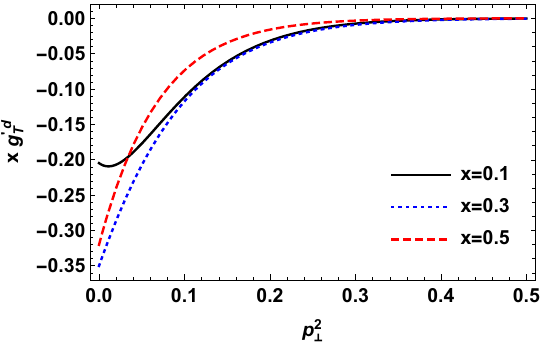}
			\hspace{0.05cm}
			(c)\includegraphics[width=7.5cm]{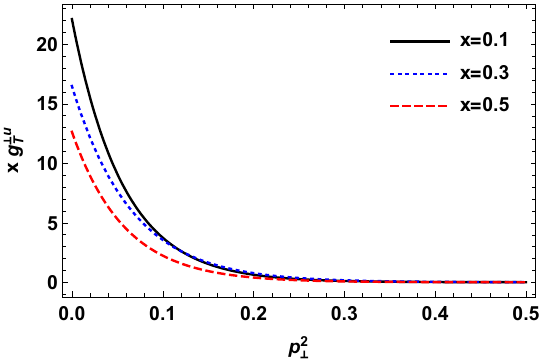}
			\hspace{0.05cm}
			(d)\includegraphics[width=7.5cm]{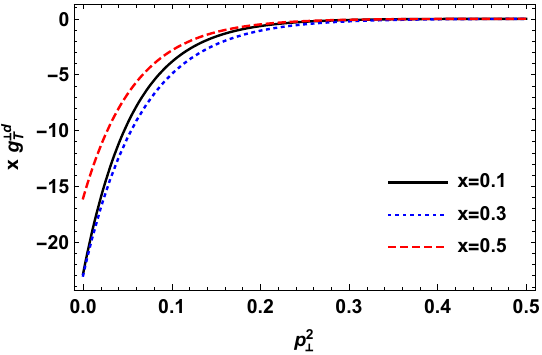}
			\hspace{0.05cm}
			(e)\includegraphics[width=7.5cm]{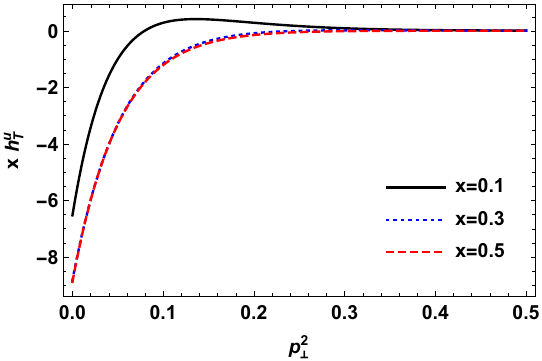}
			\hspace{0.05cm}
			(f)\includegraphics[width=7.5cm]{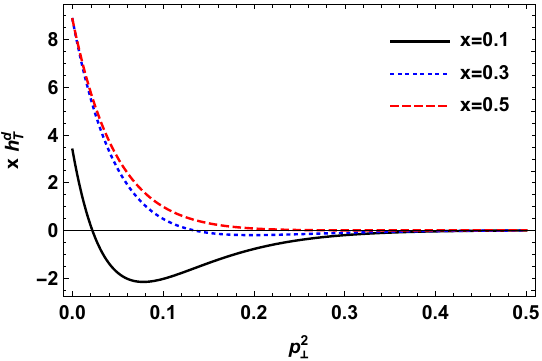}
			\hspace{0.05cm}
			(g)\includegraphics[width=7.48cm]{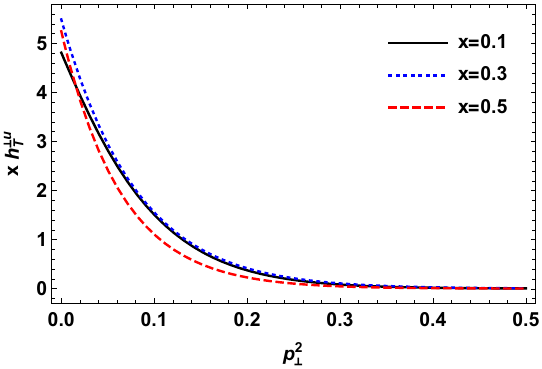}
			\hspace{0.05cm}
			(h)\includegraphics[width=7.48cm]{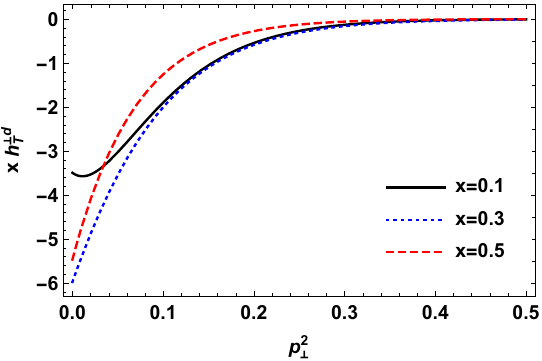}
			\hspace{0.05cm}\\
		\end{minipage}
		\caption{\label{figTMDvp3} (Color online) The transversely polarized TMDs $x g_{T}^{'\nu}(x, {\bf p_\perp^2}), x g_{T}^{\perp\nu}(x, {\bf p_\perp^2}),~ x h_{T}^{\nu}(x, {\bf p_\perp^2})$ and $ x h_{T}^{\perp\nu}(x, {\bf p_\perp^2})$  plotted with respect to ${\bf p_\perp^2}$ at different values of $x$. The left and right column correspond to $u$ and $d$ quarks respectively.}
	\end{figure*}
	\begin{figure*}
		\centering
		\begin{minipage}[c]{0.98\textwidth}
			(a)\includegraphics[width=7.5cm,clip]{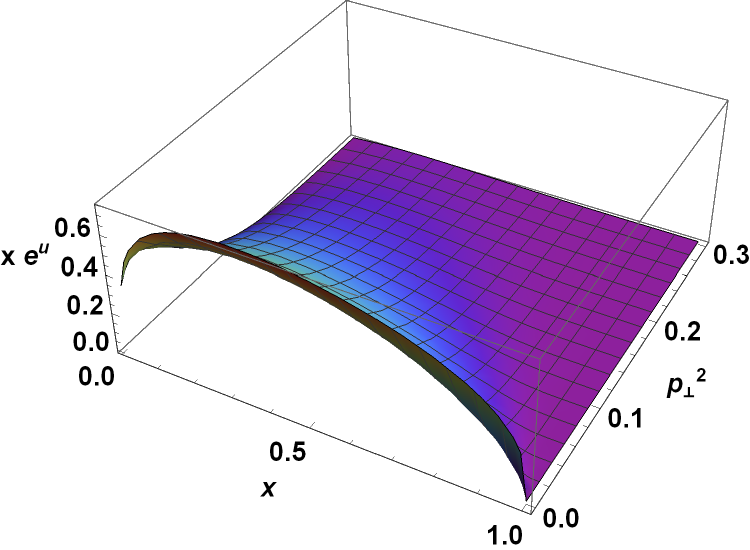}
			\hspace{0.05cm}
			(b)\includegraphics[width=7.5cm,clip]{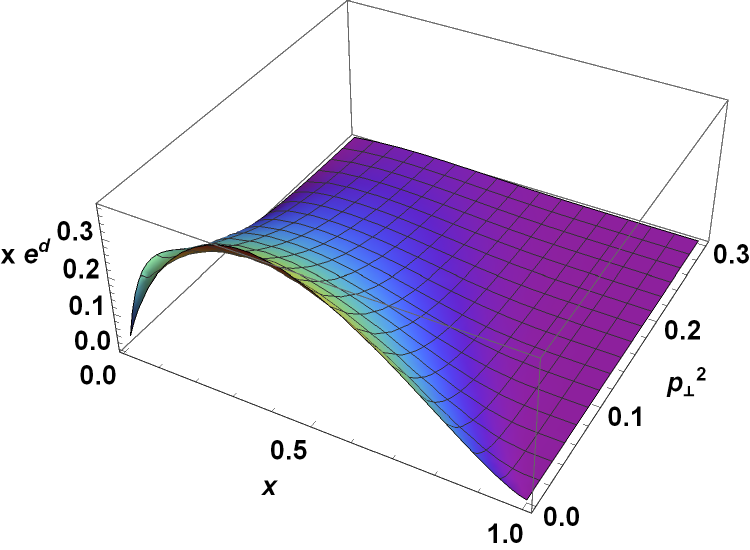}
			\hspace{0.05cm}
			(c)\includegraphics[width=7.5cm,clip]{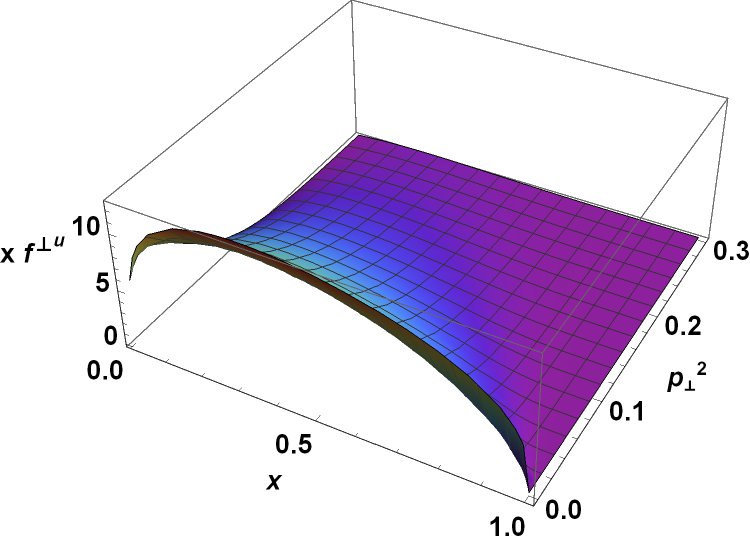}
			\hspace{0.05cm}
			(d)\includegraphics[width=7.5cm,clip]{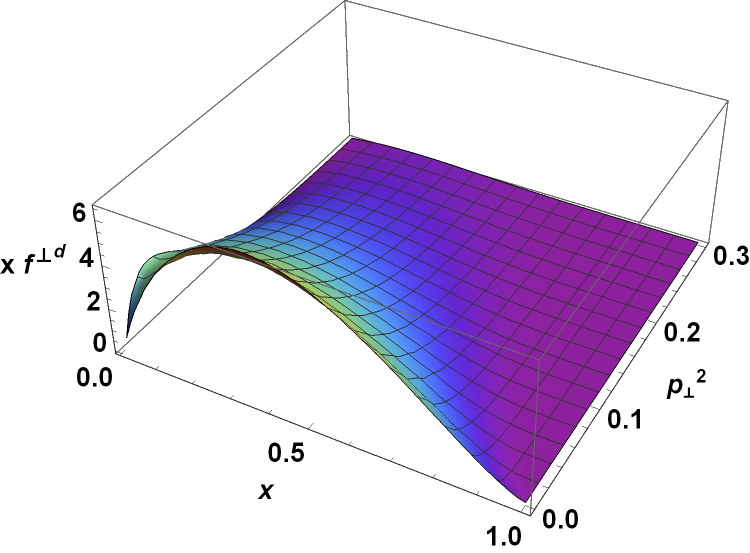}
			\hspace{0.05cm}\\
		\end{minipage}
		\caption{\label{3dfig1}(Color online) Plots of unpolarized TMDs $x e^{\nu}(x, {\bf p_\perp^2})$ and $x f^{\perp\nu}(x, {\bf p_\perp^2})$ with respect to $x$ and ${\bf p_\perp^2}$. The left and right column correspond to $u$ and $d$ quarks respectively.}
	\end{figure*}
	\begin{figure*}
		\centering
		\begin{minipage}[c]{0.98\textwidth}
			(a)\includegraphics[width=7.5cm,clip]{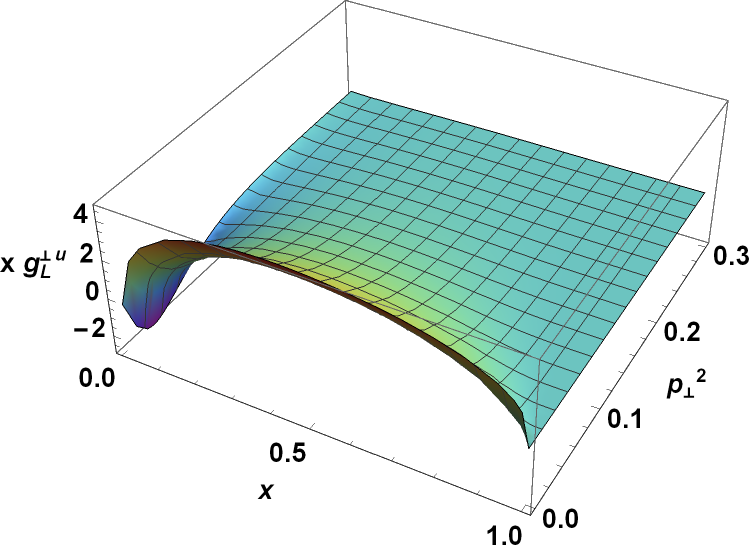}
			\hspace{0.05cm}
			(b)\includegraphics[width=7.5cm,clip]{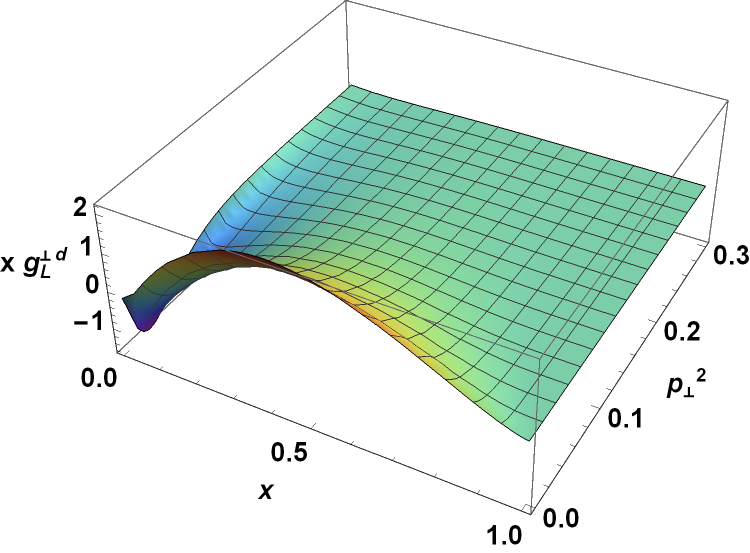}
			\hspace{0.05cm}
			(c)\includegraphics[width=7.5cm]{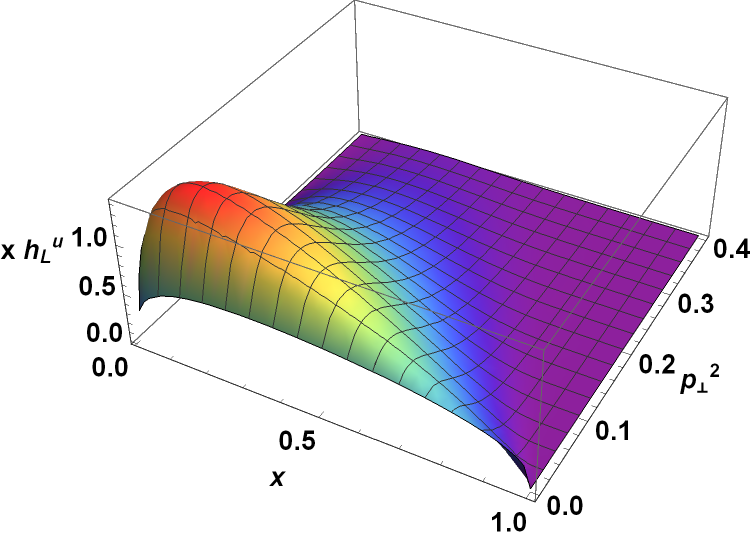}
			\hspace{0.05cm}
			(d)\includegraphics[width=7.5cm]{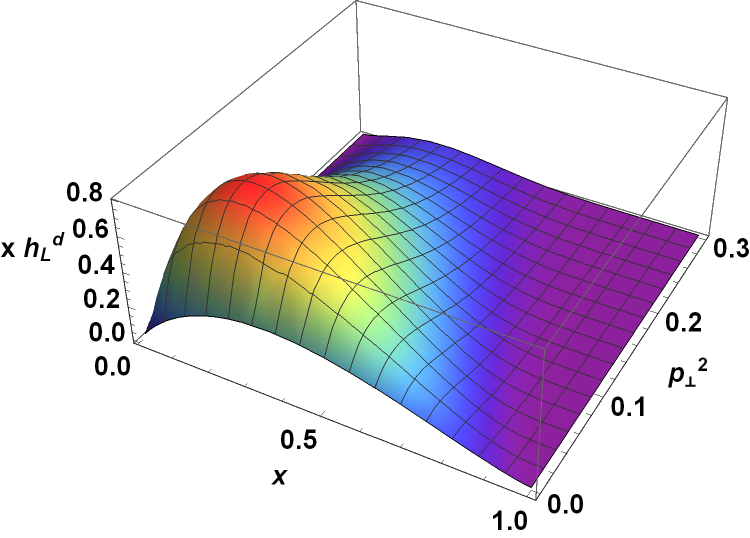}
			\hspace{0.05cm}\\
		\end{minipage}
		\caption{\label{3dfig2}(Color online) Plots of longitudinally polarized TMDs  $x g_{L}^{\perp\nu}(x, {\bf p_\perp^2})$ and $~x h_{L}^{\nu}(x, {\bf p_\perp^2})$ with respect to $x$ and ${\bf p_\perp^2}$. The left and right column correspond to $u$ and $d$ quarks respectively.}
	\end{figure*}
	\begin{figure*}
		\centering
		\begin{minipage}[c]{0.98\textwidth}
			(a)\includegraphics[width=7.2cm]{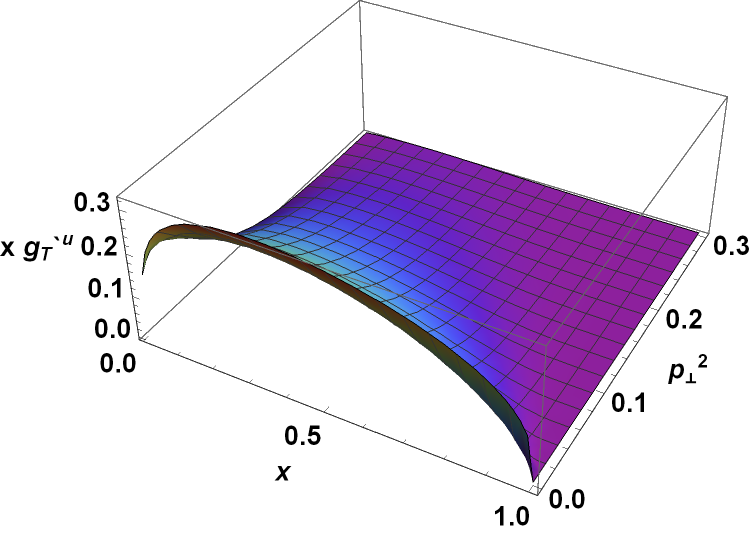}
			\hspace{0.05cm}
			(b)\includegraphics[width=7.2cm]{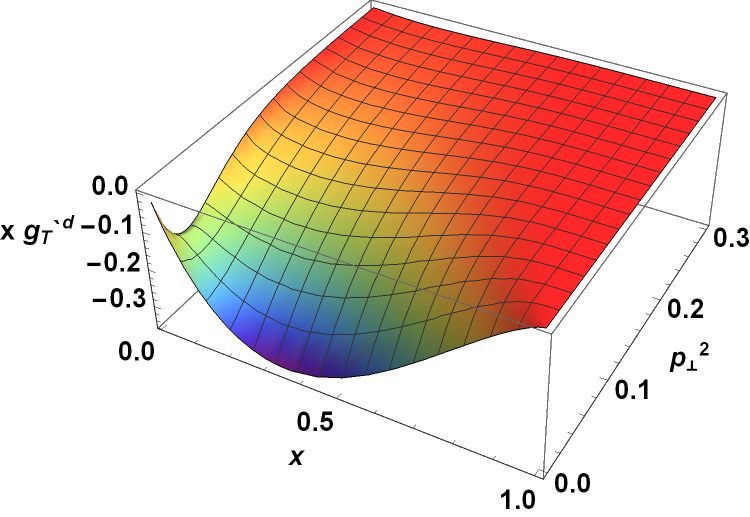}
			\hspace{0.05cm}
			(c)\includegraphics[width=7.2cm]{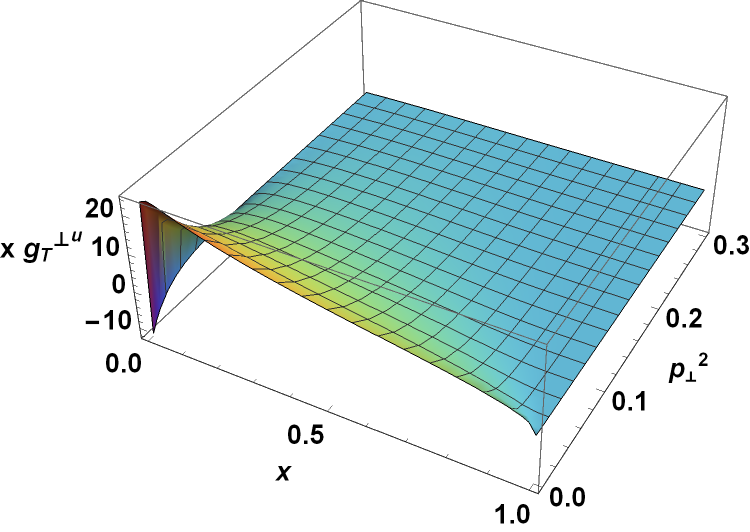}
			\hspace{0.05cm}
			(d)\includegraphics[width=7.2cm]{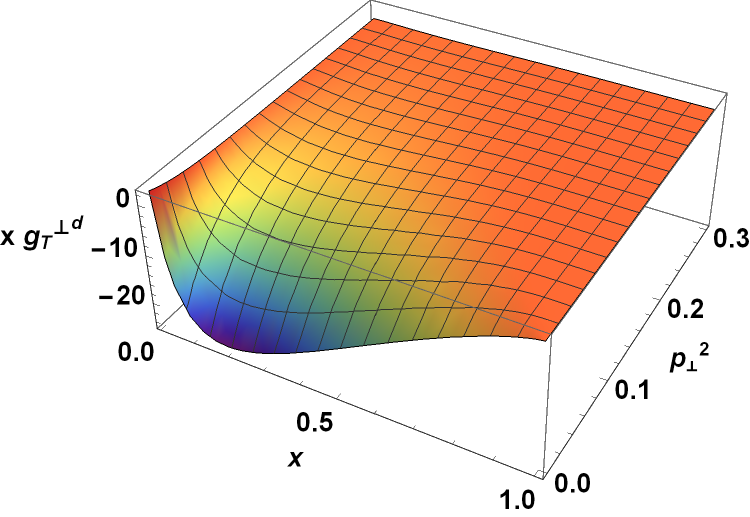}
			\hspace{0.05cm}
			(e)\includegraphics[width=7.2cm]{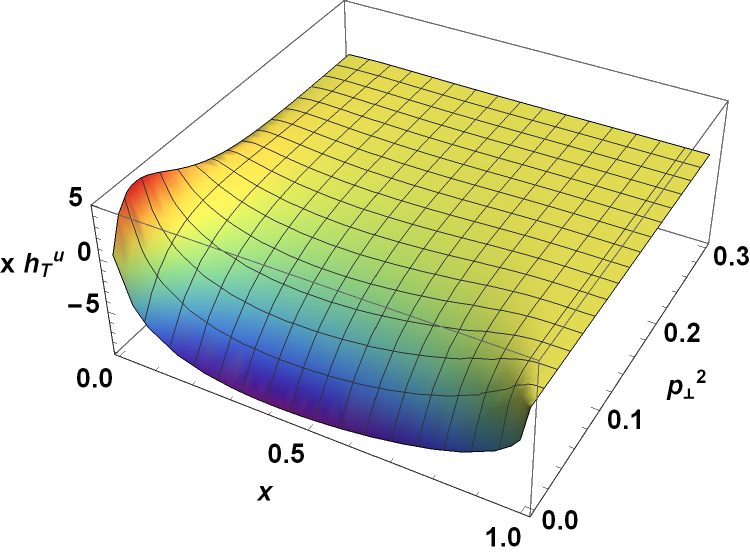}
			\hspace{0.05cm}
			(f)\includegraphics[width=7.2cm]{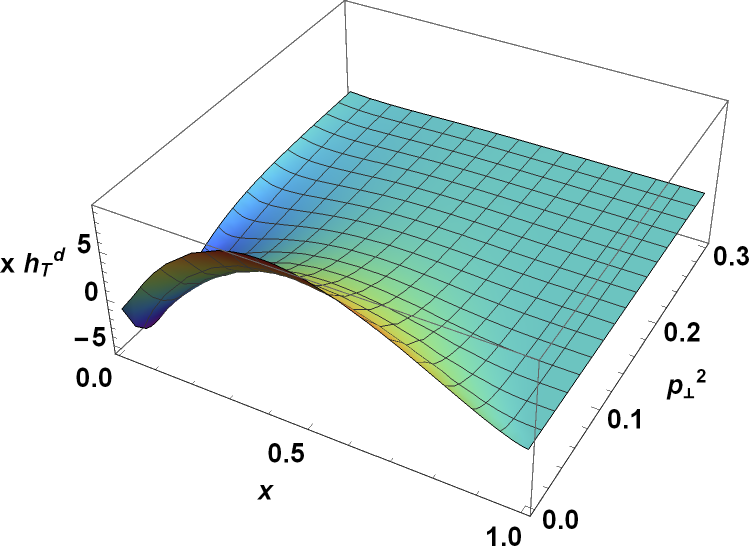}
			\hspace{0.05cm}
			(g)\includegraphics[width=7.2cm]{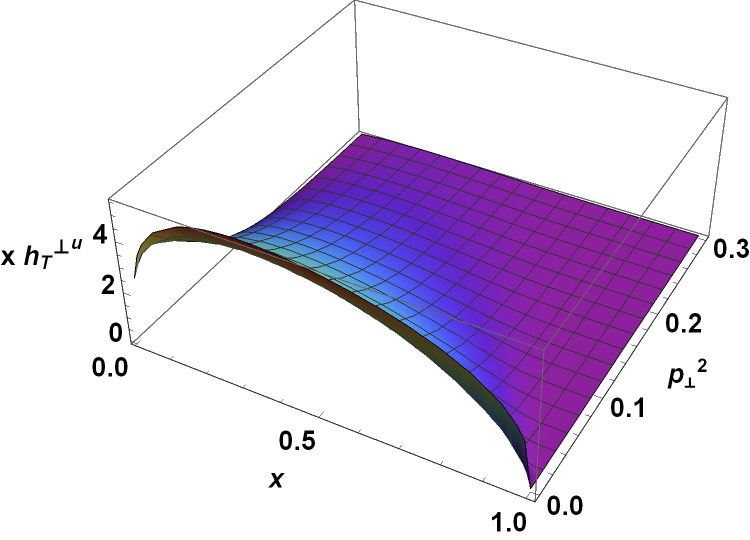}
			\hspace{0.05cm}
			(h)\includegraphics[width=7.2cm]{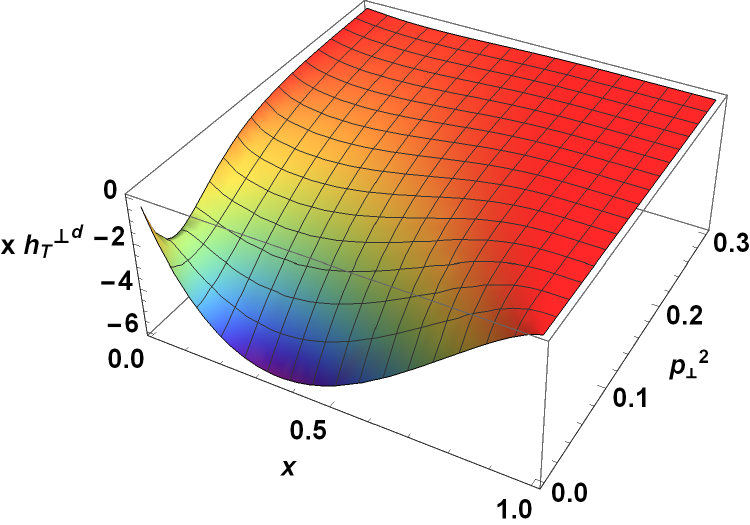}
			\hspace{0.05cm}\\
		\end{minipage}
		\caption{\label{3dfig3}(Color online) Plots of transversely polarized TMDs  $x g_{T}^{'\nu}(x, {\bf p_\perp^2}), x g_{T}^{\perp\nu}(x, {\bf p_\perp^2}),~ x h_{T}^{\nu}(x, {\bf p_\perp^2})$ and $ x h_{T}^{\perp\nu}(x, {\bf p_\perp^2})$ with respect to $x$ and ${\bf p_\perp^2}$. The left and right column correspond to $u$ and $d$ quarks respectively.}
	\end{figure*}
	The longitudinally polarized TMD $~x h_{L}^{\nu}(x, {\bf p_\perp^2})$ is plotted in Fig. \ref{figTMDvx2} (c) and \ref{figTMDvx2} (d), whose PDF $~x h_{L}^{\nu}(x)$ has been shown in Fig. \ref{Figitmd} (f) which is chirally odd and have no access through DIS. Even though not much known about this PDF phenomenologically,  $~x h_{L}^{\nu}(x)$ gives a contribution in the single spin asymmetries in SIDIS \cite{Bastami:2020rxn,Mulders:1995dh}. In CPM, a relation $ h_{L}^{\nu}(x, {\bf p_\perp^2})= 2 g_{T}^{\nu}(x, {\bf p_\perp^2})$ exists, but in present model no such relation is observed. It will be interesting to check if this relation is phenomenologically possible or not. As seen in Fig. \ref{figTMDvx3} (a) and \ref{figTMDvx3} (b), we have noticed the results for $x g_{T}^{'\nu}(x, {\bf p_\perp^2})$. It is observed that the magnitude for $u$ quark is lesser than that for $d$ quarks. At a very low value of longitudinal momentum fraction $x$, a minor fluctuation is observed in the TMD $x g_{T}^{\perp\nu}(x, {\bf p_\perp^2})$ as shown in Fig. \ref{figTMDvx3} (c) and \ref{figTMDvx3} (d). Due to flavor dependence in transversely polarized TMD, it shows a maxima for $u$ quarks and a minima for $d$ quarks. The TMD $x h_{T}^{\nu}(x, {\bf p_\perp^2})$ shows both maxima and as well as minima for both quarks flavor as shown in Fig. \ref{figTMDvx3} (e) and \ref{figTMDvx3} (f). $ x h_{T}^{\perp\nu}(x, {\bf p_\perp^2})$ is plotted in Fig. \ref{figTMDvx3} (g) and \ref{figTMDvx3} (h). With increase in $x$, the TMD first increases (decreases) and then decreases (increases) to give a maxima (minima) in the case of $u$ ($d$) quarks. \par
	
	To look clearly on the nature of transverse momentum dependence on TMDs, 
	there is a need to plot them with respect to ${\bf p_\perp^2}$ alone. In Figs. (\ref{figTMDvp1}), (\ref{figTMDvp2}) and (\ref{figTMDvp3}), these TMDs are plotted with respect to ${\bf p_\perp^2}$ at different discrete values of $x$, i.e., $x=0.1$ (black curve), $x=0.3$ (dotted blue curve) and $x=0.5$ (dashed red curve). In these figures, the left and right column correspond to $u$ and $d$ flavor of quarks respectively. It should be noted here that at the value of ${\bf p_\perp^2}$ greater than and equal to $0.2 {\ \rm GeV}^2$, the value of all TMDs are significantly negligible. In Fig. \ref{figTMDvp1}, the amplitude of these TMDs decrease to reach ${\bf p_\perp^2}$ axis with an increase in ${\bf p_\perp^2}$. In Fig. \ref{figTMDvp2} (a) and \ref{figTMDvp2} (b), with increase in ${\bf p_\perp^2}$, the amplitude of the longitudinally polarized TMD $ x g_{L}^{\perp\nu}(x, {\bf p_\perp^2})$ decreases to meet ${\bf p_\perp^2}$ axis. However, for the curve corresponding to $x=0.1$, the TMD first decreases and then increases to meet the horizontal axis. In the plot of longitudinally polarized TMD $x h_{L}^{\nu}(x, {\bf p_\perp^2})$ for both the flavors, with increase in transverse momentum ${\bf p_\perp^2}$, the magnitude of TMD first increases and then decreases to give a maxima as shown in Fig. \ref{figTMDvp2} (c) and \ref{figTMDvp2} (d). In Fig. \ref{figTMDvp3}, for all the TMDs, with increase in ${\bf p_\perp^2}$, the amplitude of TMD decreases to reach ${\bf p_\perp^2}$ axis. It should be noted that there are some exceptions for $d$ quarks in this trend and these exceptions corresponds to the curve of longitudinal momentum fraction value $x=0.1$.
	%
	%
	
	In order to study the dependence of TMDs 
	with simultaneous change in variables $x$ and ${\bf p_\perp^2}$, we have plotted their 3-D variation. In Figs. (\ref{3dfig1}), (\ref{3dfig2}) and (\ref{3dfig3}), these TMDs are plotted for both $u$ and $d$ quarks. In all these figures left column is for $u$  quarks and right is for $d$ quarks. Let us start our discussion with unpolarized TMDs $x e^{\nu}(x, {\bf p_\perp^2})$ and $x f^{\perp\nu}(x, {\bf p_\perp^2})$, which are plotted in Fig. \ref{3dfig1} (a) to \ref{3dfig1} (d). We observe that they have positive peaks for both the quark flavors as expected since they are directly related to the leading twist unpolarized TMD $f_1^q(x, {\bf p_\perp})$ via Eqs. \eqref{ec} and \eqref{fpc}. In Figs. \ref{3dfig2} (a) and \ref{3dfig2} (b), longitudinally polarized TMD $x g_{L}^{\perp\nu}(x, {\bf p_\perp^2})$ is plotted having a range of both positive and negative values for both $u$ and $d$ quarks. The longitudinally polarized TMD $~x h_{L}^{\nu}(x, {\bf p_\perp^2})$ is plotted in Fig. \ref{3dfig2} (c) and \ref{3dfig2} (d) and it has positive peaks for both $u$ and $d$ quarks. In Fig. \ref{3dfig3} (a) and \ref{3dfig3} (b), the TMD $~x g_{T}^{'\nu}(x, {\bf p_\perp^2})$ is plotted and it remains positive for $u$ quarks and negative for $d$ quarks over the whole domain. The TMD $~x g_{T}^{\perp\nu}(x, {\bf p_\perp^2})$ is plotted in Fig. \ref{3dfig3} (c) and \ref{3dfig3} (d). Except for the variation in the very low $x$ values, $~x g_{T}^{\perp\nu}(x, {\bf p_\perp^2})$ is  positive for $u$ quarks and negative for $d$ quarks. The transversely polarized TMD $ x h_{T}^{\nu}(x, {\bf p_\perp^2})$ is plotted for $u$ and $d$ quarks in Fig. \ref{3dfig3} (e) and \ref{3dfig3} (f) respectively. For both flavors, the variation ranges to both positive and negative values. In Fig. \ref{3dfig3} (g) and \ref{3dfig3} (h), $ x h_{T}^{\perp\nu}(x, {\bf p_\perp^2})$ TMDs is plotted for $u$ and $d$ quarks. It is observed that $ x h_{T}^{\perp\nu}(x, {\bf p_\perp^2})$ is positive for $u$ quarks and negative for $d$ quarks.

	\subsection{Relation between leading and sub-leading twist TMDs}\label{secrelation}
	It is important to discuss here that, in the present model, certain relations exist between the leading  and sub-leading twist TMDs. These relations  are important in order to check the consistency of the model results and can be derived by using the QCD equation of motion. These relations can be expressed as
	\begin{eqnarray}
		xe^q(x, {\bf p_\perp})           &=& x\tilde{e}^q(x, {\bf p_\perp}) + \frac{m}{M}\,f_1^q(x, {\bf p_\perp}),
		\label{ec} \\
		xf^{\perp q}(x, {\bf p_\perp})   &=& x\tilde{f}^{\perp q}(x, {\bf p_\perp})+ f_1^q(x, {\bf p_\perp}),\phantom{\frac{1}{1}} \label{fpc} \\
		xg_L^{\perp q}(x, {\bf p_\perp}) &=& x\tilde{g}_L^{\perp q}(x, {\bf p_\perp}) + g_1^q(x, {\bf p_\perp})+\frac{m}{M}\,h_{1L}^{\perp q}(x, {\bf p_\perp}),\label{glpc}\\
		xg_T^{\perp q}(x, {\bf p_\perp}) &=& x\tilde{g}_T^{\perp q}(x, {\bf p_\perp})+g_{1T}^{\perp q}(x, {\bf p_\perp}) +\frac{m}{M}\,h_{1T}^{\perp q}(x, {\bf p_\perp}),\label{gtpc}\\
		xg_T^q(x, {\bf p_\perp})        &=& x\tilde{g}_T^q(x, {\bf p_\perp}) +\frac{\vec{p}_T^{\:2}}{2M^2}\,g_{1T}^{\perp q}(x, {\bf p_\perp})+\frac{m}{M}\,h_1^q(x, {\bf p_\perp}),\label{gtc}\\
		xh_L^q(x, {\bf p_\perp})         &=& x\tilde{h}_L^q(x, {\bf p_\perp}) -\frac{\vec{p}_T^{\:2}}{M^2}\,h_{1L}^{\perp q}(x, {\bf p_\perp})+\frac{m}{M}\,g_1^q(x, {\bf p_\perp}),\label{hlc}\\
		xh_T^q(x, {\bf p_\perp})         &=& x\tilde{h}_T^q(x, {\bf p_\perp}) - h_1^q(x, {\bf p_\perp}) -\frac{\vec{p}_T^{\:2}}{2 M^2}\, h_{1T}^{\perp q}(x, {\bf p_\perp})
		+\frac{m}{M}\,g_{1T}^{\perp q}(x, {\bf p_\perp}),\label{htc}\\
		xh_T^{\perp q}(x, {\bf p_\perp}) &=& x\tilde{h}_T^{\perp q}(x, {\bf p_\perp})+h_1^q(x, {\bf p_\perp})-\frac{\vec{p}_T^{\:2}}{2 M^2}\, h_{1T}^{\perp q}(x, {\bf p_\perp}),\label{htpc}\\
		xg_T^{\prime q}(x, {\bf p_\perp})&=& x\tilde{g}_T^{\prime q}(x, {\bf p_\perp}) + \frac{m}{M}h_1^q(x, {\bf p_\perp}) - \frac{m}{M}\,
		\frac{\vec{p}_T^{\:2}}{2 M^2}\, h_{1T}^{\perp q}(x, {\bf p_\perp})\label{gtprc},
	\end{eqnarray}
	where the TMD $g_{T}^{ \nu} (x,\textbf{p}_{\perp}^2)$ can be obtained from $g_{T}^{' \nu} (x,\textbf{p}_{\perp}^2)$ and $g_{T}^{\perp  \nu} (x,\textbf{p}_{\perp}^2)$ as
	\begin{eqnarray}
		g^{\nu }_T (x,\textbf{p}_{\perp}^2) &=& g^{'\nu }_{T} (x,\textbf{p}_{\perp}^2) + \frac{\textbf{p}^2_{\perp}}{2M^2}g_{T}^{\perp  \nu} (x,\textbf{p}_{\perp}^2).\label{gtconv}
	\end{eqnarray}
	The tilde functions in above equations  are used to represent the genuine twist-3 contributions and they arise from  the quark-gluon correlator. After rewriting the quark-quark correlator in Eq. \eqref{TMDcor} for the extraction of leading \cite{Maji:2017bcz} and sub-leading twist TMDs,  it is found that all tilde terms vanish due to the absence of quark-model interactions in LFQDM.

	%
	
	\subsection{Average Transverse Momentum $\langle\bfp^r(w)\rangle^\nu$}\label{secavgtr}
	The average transverse momenta ($ r=1 $) and the average square transverse momenta ($ r=2 $) for TMD $ w^{\nu}(x,\bfp^2) $ in the LFQDM can be defined as
	\be
	\langle p_\perp^r(w)\rangle^\nu= \frac{\int dx\int d^2p_\perp p^r_\perp w^{\nu}(x,\bfp^2)}{\int dx\int d^2p_\perp w^{\nu}(x,\bfp^2)}.
	\label{Eqavg}
	\ee
	\begin{table}[h]
		\centering 
		\begin{tabular}{||c||c|c|c|c|c|c|c|c|c|c||}
			\hline
			\hline
			\text{TMD w}~~&~~$ \nu  $~~&~~$ e^{\nu} $~~&~~$ f^{\perp\nu} $~~&~~$ g_{L}^{\perp\nu} $~~&~~$ g_{T}^{'\nu} $~~&~~$ g_{T}^{\perp\nu} $~~&~~$ h_{L}^{\nu} $~~&~~$ h_{T}^{\nu} $~~&~~$ h_{T}^{\perp\nu} $~~&~~$ g_{T}^{\nu} $\\
			\hline
			\hline
			\text{LFQDM}~~&~~$ \langle p_\perp \rangle^{u} $~~&~~$ 0.99 $~~&~~$ 0.99 $~~&~~$ 1.25 $~~&~~$ 0.99 $~~&~~$ 0.88 $~~&~~$ 1.23 $~~&~~$ 1.25 $~~&~~$ 0.99 $~~&~~$ 1.17 $\\
			\hline
			\text{LFQDM}~~&~~$ \langle p_\perp \rangle^{d} $~~&~~$ 1.00 $~~&~~$ 1.00 $~~&~~$ 1.13 $~~&~~$ 1.00 $~~&~~$ 1.07 $~~&~~$ 1.25 $~~&~~$ 1.13 $~~&~~$ 1.00 $~~&~~$ 1.18 $\\
			\hline
			\text{LFCQM}~~&~~$ \langle p_\perp \rangle^{\nu} $~~&~~$ 0.92 $~~&~~$ 0.92 $~~&~~$ \text{$....$} $~~&~~$ \text{$....$} $~~&~~$ \text{$....$} $~~&~~$ \text{$....$} $~~&~~$ \text{$....$} $~~&~~$ \text{$....$} $~~&~~$ \text{$....$} $ \\
			\hline
			\hline
		\end{tabular}
		\caption{Average transverse momentum for sub-leading twist T-even TMDs in our model (LFQDM) and in LFCQM \cite{Lorce:2014hxa}.}
		\label{tab_avgP} 
	\end{table}
	
	\begin{table}[h]
		\centering 
		\begin{tabular}{||c||c|c|c|c|c|c|c|c|c|c||}
			\hline
			\hline
			\text{TMD w}~~&~~$ \nu  $~~&~~$ e^{\nu} $~~&~~$ f^{\perp\nu} $~~&~~$ g_{L}^{\perp\nu} $~~&~~$ g_{T}^{'\nu} $~~&~~$ g_{T}^{\perp\nu} $~~&~~$ h_{L}^{\nu} $~~&~~$ h_{T}^{\nu} $~~&~~$ h_{T}^{\perp\nu} $~~&~~$ g_{T}^{\nu} $ \\
			\hline
			\text{LFQDM}~~&~~$ \langle p_\perp^2 \rangle^{u} $~~&~~$ 0.94 $~~&~~$ 0.94 $~~&~~$ 1.30 $~~&~~$ 0.94 $~~&~~$ 0.77 $~~&~~$ 1.39 $~~&~~$ 1.30 $~~&~~$ 0.94 $~~&~~$ 1.28 $\\
			\hline
			\text{LFQDM}~~&~~$ \langle p_\perp^2 \rangle^{d} $~~&~~$ 0.96 $~~&~~$ 0.96 $~~&~~$ 1.15 $~~&~~$ 0.96 $~~&~~$ 1.06 $~~&~~$ 1.43 $~~&~~$ 1.15 $~~&~~$ 0.96 $~~&~~$ 1.30 $ \\
			\hline
			\text{LFCQM}~~&~~$ \langle p_\perp^2 \rangle^{\nu} $~~&~~$ 0.86 $~~&~~$ 0.86 $~~&~~$ \text{$....$} $~~&~~$ \text{$....$} $~~&~~$ \text{$....$} $~~&~~$ \text{$....$} $~~&~~$ \text{$....$} $~~&~~$ \text{$....$} $~~&~~$ \text{$....$} $ \\
			\hline
			\text{Bag Model}~~&~~$ \langle p_\perp^2(x_v) \rangle_{Gauss} $~~&~~$ 0.68 $~~&~~$ 0.94 $~~&~~$ 1.11 $~~&~~$ \text{$....$} $~~&~~$ 1.11 $~~&~~$ 1.01 $~~&~~$ 1.11 $~~&~~$ 0.94 $~~&~~$ 0.84 $ \\
			\hline
			\hline
		\end{tabular}
		\caption{Average transverse momentum squares in sub-leading twist T-even TMDs from our model (LFQDM), LFCQM \cite{Lorce:2014hxa} and the bag model \cite{Avakian:2010br}.}
		\label{tab_avgP2} 
	\end{table}
	
	The results of average transverse momenta for sub-leading twist T-even TMDs in the LFQDM have been shown in Table \ref{tab_avgP}. We have also tabulated the values from the LFCQM for comparison \cite{Lorce:2014hxa} in the table. The $\langle p_\perp \rangle$ in the LFCQM model have also been defined according to (\ref{Eqavg}). All results are expressed in units of their corresponding $f_1^{\nu}$ value, which is $ \langle p_\perp \rangle^{u} =0.23\,{\rm GeV}$, $\langle p_\perp \rangle^{d}=0.24\,{\rm GeV}$ in our model and $ \langle p_\perp \rangle^{\nu} =0.24\,{\rm GeV} $ in the LFCQM \cite{Lorce:2014hxa}.  
	
	It is beneficial to compare our results with the available results of the bag model \cite{Avakian:2010br}, however, the average transverse momentum $\langle p_\perp^1(w)\rangle^\nu$ and average square transverse momentum $\langle p_\perp^2(w)\rangle^\nu$ of the TMDs defined in Eq.~(\ref{Eqavg}) may not exist in the bag model, because their momentum-space wave-function components $t_i(k)$, do not vanish sufficiently fast at large value of momentum. The Gaussian widths expressed as
	\be\label{Eq:Gauss-width}
	\langle p_\perp^2(x)\rangle_{\rm Gauss} = \pi \;\frac{w^{\nu}(x,0)}{w^{\nu}(x)},
	\ee
	seems rather promising for the comparison \cite{Avakian:2010br}. Average transverse momentum squares in sub-leading twist T-even TMDs from our model (LFQDM), LFCQM \cite{Lorce:2014hxa} and the bag model \cite{Avakian:2010br} are tabulated  in Table \ref{tab_avgP2}. The $\langle p_\perp^2 \rangle$ in the LFQDM and LFCQM are defined by Eq. (\ref{Eqavg}). The findings of the bag model for the Gaussian widths are determined according to Eq. (\ref{Eq:Gauss-width}) and measured at the valence-$x$ point $x_v=0.3$ due to the weak $x$-dependence of the $\langle p_\perp^2(x)\rangle_{\rm Gauss}$. The results are listed in terms of the value for $f_1^{\nu}$ in that model, which is $\langle p_\perp^2\rangle^{u} =0.066~{\rm GeV}^2$ and $\langle p_\perp^2 \rangle^{d}=0.075~{\rm GeV}^2$ instance of LFQDM, $  \langle p_\perp^2 \rangle^{\nu}=0.080~{\rm GeV}^2 $ in LFCQM \cite{Lorce:2014hxa} and $\langle p_\perp^2(x_v)\rangle_{\rm Gauss}^{(f_1)}=0.077~{\rm GeV}^2$ in the case of the bag model \cite{Avakian:2010br}. In Ref. \cite{Avakian:2010br}, the value of $\langle p_\perp^2 \rangle$ for TMD $ g_{T}^{'\nu} $ is not given, rather, the value of $\langle p_\perp^2 \rangle$ for TMD $ g_{T}^{\nu} $ is provided. Therefore, by using Eq. \eqref{gtconv}, we have calculated and tabulated $\langle p_\perp^2 \rangle$ values for both of them in our model. 
	
	%
	%
	\begin{figure*}
		\centering
		\begin{minipage}[c]{0.98\textwidth}
			(a)\includegraphics[width=7.5cm,clip]{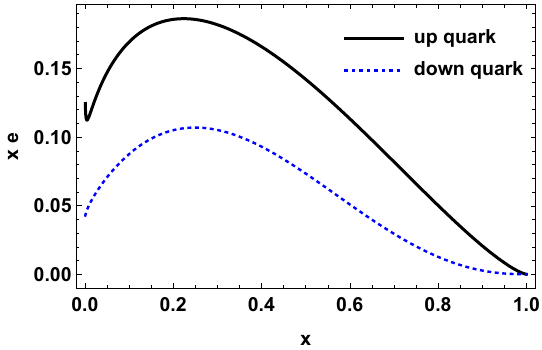}
			\hspace{0.05cm}
			(b)\includegraphics[width=7.5cm,clip]{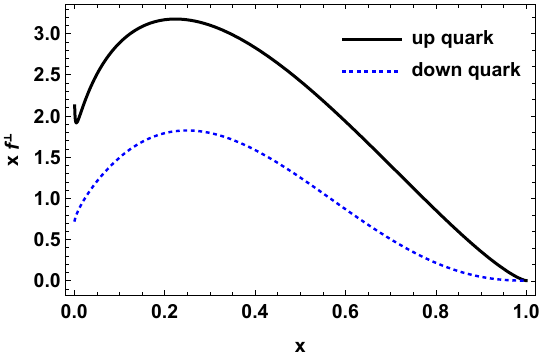}
			\hspace{0.05cm}
			(c)\includegraphics[width=7.5cm,clip]{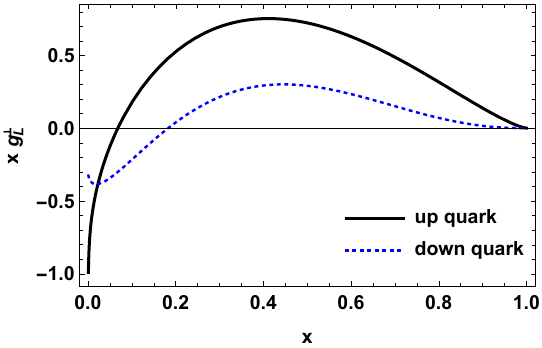}
			\hspace{0.05cm}
			(d)\includegraphics[width=7.5cm,clip]{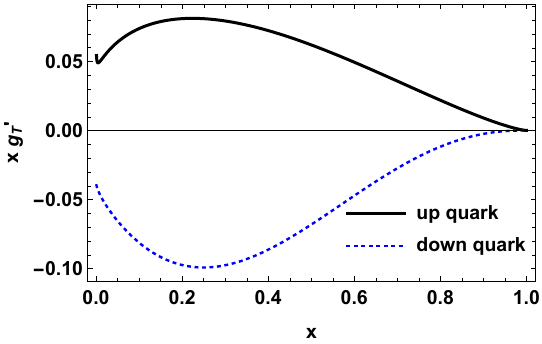}
			\hspace{0.05cm}
			(e)\includegraphics[width=7.5cm,clip]{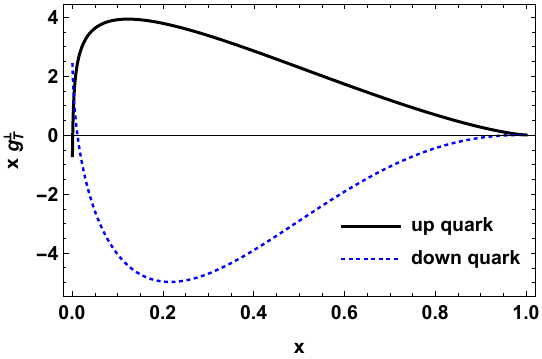}
			\hspace{0.05cm}
			(f)\includegraphics[width=7.5cm,clip]{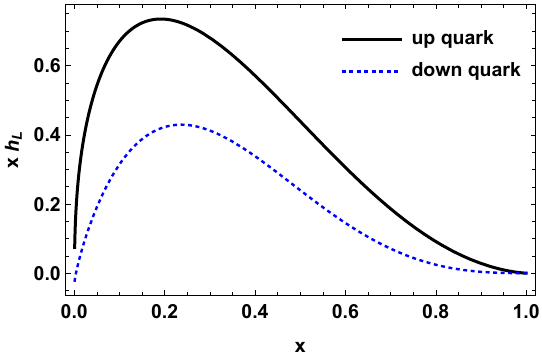}
			\hspace{0.05cm}
			(g)\includegraphics[width=7.48cm,clip]{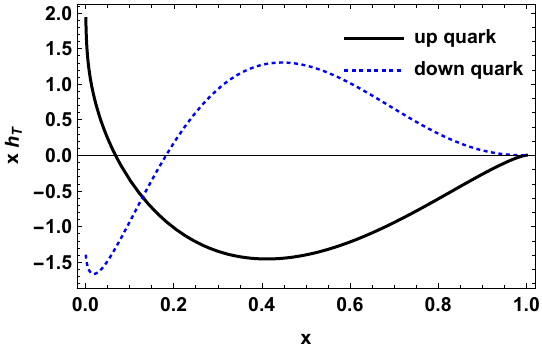}
			\hspace{0.05cm}
			(h)\includegraphics[width=7.48cm,clip]{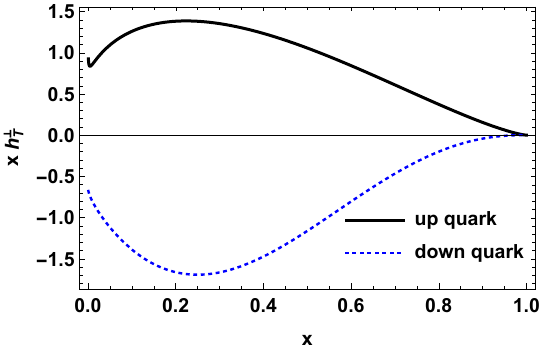}
			\hspace{0.05cm}\\
		\end{minipage}
		\caption{\label{Figitmd} (Color online) Plots of $x e^{\nu}(x),~x f^{\perp\nu}(x),~x g_{L}^{\perp\nu}(x),~x g_{T}^{'\nu}(x),~x g_{T}^{\perp\nu}(x),~x h_{L}^{\nu}(x),~x h_{T}^{\nu}(x)$ and $x h_{T}^{\perp\nu}(x)$ PDFs  plotted with respect to $x$. Black and dotted blue curve correspond to $u$ and $d$ quarks respectively.}
	\end{figure*}
	\subsection{Parton Distribution Functions}\label{secitmd}
	To have a better understanding of the sub-leading twist TMDs, we have derived their PDFs by integrating the TMDs over the quark's transverse momentum ${\bf p_\perp}$. PDFs obtained here at the model initial scale and model dependent relations among them are valid only at tree level. In Fig. \ref{Figitmd}, the PDFs  $x e^{\nu}(x),~ x f^{\perp\nu}(x)$ $~x vg_{L}^{\perp\nu}(x),~x g_{T}^{'\nu}(x),~x g_{T}^{\perp\nu}(x),~x h_{L}^{\nu}(x),~x h_{T}^{\nu}(x)$ and $ x h_{T}^{\perp\nu}(x)$ have been plotted with respect to longitudinal momentum fraction $x$. In this figure, the black and dotted blue curves correspond to $u$ and $d$ quarks respectively. Keeping the magnitude aside in Fig. \ref{Figitmd}, it is observed that for the unpolarized \bigg($x e^{\nu}(x)$ and $x f^{\perp\nu}(x)$\bigg) and the longitudinally polarized $\bigg(x g_{L}^{\perp\nu}(x)$ and $x h_{L}^{\nu}(x)$\bigg) PDFs no flip in sign is observed while interchanging the flavor between $u$ and $d$ quarks. However, for transversely polarized \bigg($x g_{T}^{'\nu}(x),~x g_{T}^{\perp\nu}(x), x h_{T}^{\nu}(x)$ and $ x h_{T}^{\perp\nu}(x)$\bigg) PDFs, flip in sign is observed. Let us start our discussion with unpolarized functions $x e^{\nu}(x)$ and $f^{\perp\nu}(x)$. In Fig. \ref{Figitmd} (a), the function $x e^{\nu}(x)$ is plotted which is related to unpolarized PDF $f_1^{\nu}(x)$ with a constant multiple $\frac{m}{M}$ (see Eq. \eqref{ec}). It is observed that for $x e^{\nu}(x)$ the magnitude of $u$ quarks is larger than for the $d$ quarks. $f^{\perp\nu}(x)$ is plotted in Fig. \ref{Figitmd} (b), which is equal to unpolarized PDF $f_1^{\nu}(x)$ via Eq. \eqref{fpc}. Also here the magnitude of $u$ quarks is larger than for the $d$ quarks. In Fig. \ref{Figitmd} (c), the longitudinally polarized PDF $x g_{L}^{\perp\nu}(x)$ have been plotted. For $u$ quarks, with increase in $x$ the function first increases and then decreases to meet the horizontal axis. For $d$ quarks, a similar plot is obtained except the fact that there is a minima at a very low value of longitudinal momentum fraction. PDFs $x g_{T}^{'\nu}(x)$ and $x g_{T}^{\perp\nu}(x)$ have been plotted in Fig. \ref{Figitmd} (d) and  \ref{Figitmd} (e). For these PDFs, with an increase in the longitudinal momentum fraction $x$, the functions first increase (decrease) for $u$ ($d$) quarks and then decrease (increase) to represent a maxima (minima) in between. In Fig. \ref{Figitmd} (f), the longitudinally polarized PDF $x h_{L}^{\nu}(x)$ is plotted. With increase in the longitudinal momentum fraction $x$, the function first increases to reach a maxima and then decreases for both $u$ and $d$ quarks. In this plot, the curve of $u$ quarks has a larger magnitude as compared to that  of the $d$ quarks. We have plotted the transversely polarized PDF $x h_{T}^{\nu}(x)$ in Fig. \ref{Figitmd} (g). For $u$ quarks, with an increase in $x$, the function falls from positive region to the negative region and thereafter increases to reach $x$ axis. For $d$ quarks, with an increase in $x$ the PDF first decreases and then increases to give a minima at low $x$ and after that it decreases to give a maxima. With increase in the longitudinal momentum fraction $x$, the transversely polarized PDF $ x h_{T}^{\perp\nu}(x)$ increases (decreases) to reach maxima (minima) and then decreases (increases) to meet the axis of longitudinal momentum fraction, as shown in Fig. \ref{Figitmd} (h).
	
	\section{Comparison with phenomenology}
\begin{figure*}
		\centering
		\begin{minipage}[c]{0.98\textwidth}
			(a)\includegraphics[width=7.5cm]{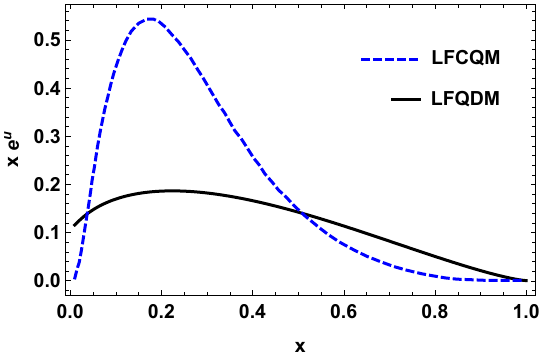}
			\hspace{0.05cm}
			(b)\includegraphics[width=7.5cm]{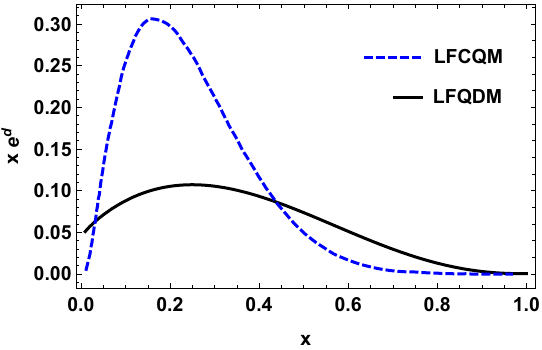}
			\hspace{0.05cm}\\
		\end{minipage}
		\caption{(Color online) $x e^{u}(x)$ and $x e^{d}(x)$ (mass-term contribution only) plotted with respect to $x$ for LFQDM and LFCQM \cite{Rodini:2019ktv}. The left and right column correspond to $u$ and $d$ quarks respectively.}
		\label{mass-term-compare}
	\end{figure*}

{\label{pheno}}
	In this section, we compare our result for $e(x)$ PDF with the recently reported CLAS collaboration results in the Wandzura-Wilczek approximation.  The best channel to access the $e(x)$ is the dihadron-SIDIS process. Recently CLAS and CLAS12 collaboration reported the result for measurement of a beam spin asymmetry from di-hadron SIDIS process  $l(p_1) + N (P) \rightarrow  l(p_2) + h_1(P_1) + h_2(P_2) + X$ \cite{Courtoy:2014ixa,Courtoy:2022kca}. CLAS data provide a good access to scalar PDF $e(x)$ by means of fragmentation functions (FFs) at both twist-2 and twist-3 level. The twist-2 dihadron FF are the unpolarized $D_1$ and chiral-odd, T-odd $H_1^\measuredangle$ where $D_1$ describe the hadronization of a quark into an unpolarized hadron pairs averaging over the quark polarization and $H_1^\measuredangle$ describe the correlation between the transverse polarization of the fragmenting quark and the azimuthal orientation of plane containing the momentum of detected hadron pair.  In Figure \ref{mass-term-compare}, we have compared our results with the mass term contributions for the scalar PDF  $e(x)$ only with the results from the LFCQM \cite{Pasquini:2018oyz}. The magnitude of $e(x)$ for both up and down quarks in LFCQM is quite large as compared to our results. This may be due to fact our model scale is quite low as compared to the LFCQM scale. 
	\begin{figure*}
		\centering
		\begin{minipage}[c]{0.98\textwidth}
			(a)\includegraphics[width=7.5cm]{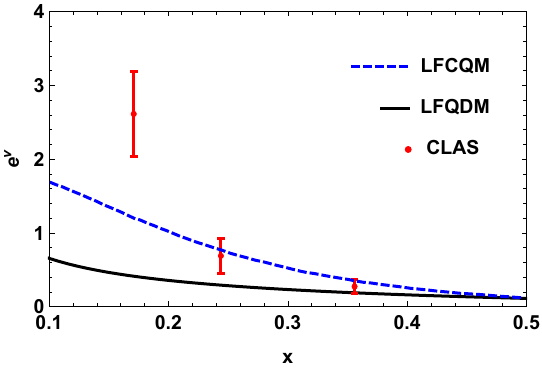}
			\hspace{0.05cm}
			(b)\includegraphics[width=7.5cm]{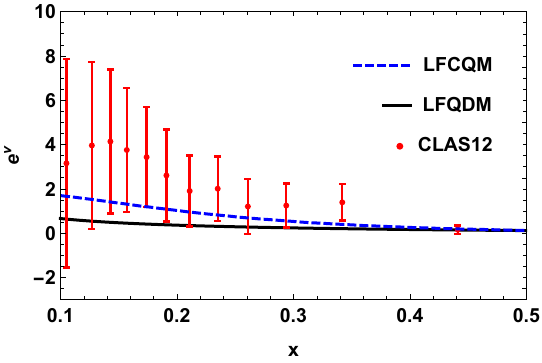}
			\hspace{0.05cm}\\
		\end{minipage}
		\caption{(Color online) Flavor combination $e^v(x)$ (mass-term contribution only) plotted with respect to $x$ in LFQDM and LFCQM \cite{Rodini:2019ktv} for CLAS \cite{Courtoy:2014ixa} and CLAS12 \cite{Courtoy:2022kca} extracted data respectively.}
		\label{clas-compare}
	\end{figure*}
	Recently, Courtoy {\it et. al.} \cite{Courtoy:2014ixa}  extracted $e(x)$ from the beam spin asymmetry data from CLAS collaboration. In Fig. \ref{clas-compare}(a) and (b) we have compare our result for the flavor combination $e^v(x) = \frac{4}{9} e^u(x) - \frac{1}{9} e^d(x)$ with the old CLAS data and updated CLAS12 data. We would like to emphasize here that this comparison corresponds to only WW approximation and  shows that model results are in good agreement at higher values of $x$. However, at lower values of $x$ the results differ in magnitude. The model results can further be improved by considering the contribution from the quark-gluon interaction term for $e(x)$.
	
	\section{Conclusion}\label{seccon}
	In this work, we have investigated the sub-leading twist T-even TMDs using systematic computations within the context of the LFQDM. Utilizing the unintegrated SIDIS quark-quark correlator, we were able to get the overlap form of TMDs. We have generated the  explicit expressions of the TMDs by considering the diquark to be a scalar and a vector. For both the $u$ and $d$ quarks, we have demonstrated the 2-D and 3-D variation of the TMDs with longitudinal momentum fraction $x$ and transverse momentum ${\bf p_\perp^2}$. In the plots of unpolarized $\bigg(x e^{\nu}(x, {\bf p_\perp^2})$ and $x f^{\perp\nu}(x, {\bf p_\perp^2})\bigg)$ and longitudinally polarized $\bigg(x g_{L}^{\perp\nu}(x, {\bf p_\perp^2})$ and $~x h_{L}^{\nu}(x, {\bf p_\perp^2})\bigg)$ TMDs, no sign flip is detected, however, a sign flip is found for the transversely polarized TMDs \bigg($x g_{T}^{'\nu}(x, {\bf p_\perp^2}),~x g_{T}^{\perp\nu}(x, {\bf p_\perp^2}), x h_{T}^{\nu}(x, {\bf p_\perp^2})$ and $ x h_{T}^{\perp\nu}(x, {\bf p_\perp^2})$\bigg). In general, the TMD's amplitude of $u$ quarks exceeds that of $d$ quarks. We have observed that as the transverse momentum increases, the amplitude of these TMDs decrease and become insignificant when ${\bf p_\perp^2}>0.3 {\ \rm GeV}^2$. There are however a few minor exceptions at extremely low values of ${\bf p_\perp^2}$.  The explanation comes from the  model itself as the wave functions used to illustrate our results are exponential of negative ${\bf p_\perp^2}$ and dependent on some other factors but when ${\bf p_\perp^2}$ is larger than or equal to $0.3 {\ \rm GeV}^2$, the exponentially diminishing component predominates. As our model operates at low $Q^2$,  quite sizable plots of TMDs $x e^{\nu}(x, {\bf p_\perp^2})$ and $ x f^{\perp\nu}(x, {\bf p_\perp^2})$ are observed. For $p_\perp^2=0.1$, the plots of $u$ and $d$ quarks for longitudinally polarized TMD $x g_{L}^{\perp\nu}(x, {\bf p_\perp^2})$ show a node around $x=0.12$ and $x=0.25$ sequentially, which is similar to the results from the CPM and the bag model. From the plots of $ g_{L}^{\perp\nu}(x, {\bf p_\perp^2})$ and $ h_{T}^{\nu}(x, {\bf p_\perp^2})$, we found that they show exact opposite behavior to each other for $u$ quark, but for $d$ quark they differ only by magnitude which indicates presence of some quark model symmetry. Contrary to CPM, the relations $ g_{L}^{\perp\nu}(x, {\bf p_\perp^2})= -  h_{T}^{\nu}(x, {\bf p_\perp^2})$ and $ h_{L}^{\nu}(x, {\bf p_\perp^2})= 2 g_{T}^{\nu}(x, {\bf p_\perp^2})$, in our model are not followed. In future, it would be fascinating to check whether these relations are phenomenologically possible or not. 
	\par We have obtained the PDFs by integrating the TMDs over the transverse momentum of quark ${\bf p_\perp}$ and studied their variation with respect to the longitudinal momentum fraction $x$. Sign of the plot for unpolarized and longitudinally polarized PDFs is same for both the flavors, however, the sign of transversely polarized PDFs flips when $u$ quarks are replaced by $d$ quarks. Additionally, it is noted that the trend of PDF is analogous to that of their corresponding TMD. We have calculated the average transverse momenta and average square transverse momenta of sub-leading twist T-even TMDs and compared it with the data from LFCQM and bag model. We have also found that our model is consistent and follows certain QCD relations between the leading and sub-leading twist T-even TMDs. Future studies will throw more light on the model independent relations of these TMDs. We have also compared our results for PDF $e(x)$ with LFCQM and CLAS data which, under WW approximation, are in close agreement at higher values of $x$. However, we believe that the inclusion of quark-gluon interactions in LFQDM can improve the results from phenomenology point of view. It is also a part of our future work.
	\par To summarize, the contributions of sub-leading twist T-even TMDs  and their PDFs have been the object of study in DIS experiments like HERMES and those being conducted in J-Lab. 
	In the future, it would be interesting to study SIDIS observables like transverse spin asymmetry (TSA) and Collins azimuthal asymmetry (CAA) as computing them will open the door to study the  TMD predictions for these observables.
 It would also be fascinating to study other distributions like  generalized transverse momentum dependent distributions (GTMDs), Wigner distributions and GPDs at the sub-leading and higher twist in LFQDM. This would not only have the possibility to illuminate the complicated multidimensional structure of hadrons  but also impose significant and decisive constraints in different kinematic regions. 
\section{Acknowledgement}
	Authors would like to thank Chandan Mondal for insightful discussion on phenomenology section. N.K. and H.D. would like to thank the Science \& Engineering Research Board, Department of Science and Technology, Government of India for providing Teacher Associateship Research Excellence Award, Grant No. TAR/2021/000157 for the financial support.

			\end{document}